\documentclass[sigconf,9pt,table]{acmart}

\makeatletter
\def\@ACM@checkaffil{
    \if@ACM@instpresent\else
    \ClassWarningNoLine{\@classname}{No institution present for an affiliation}%
    \fi
    \if@ACM@citypresent\else
    \ClassWarningNoLine{\@classname}{No city present for an affiliation}%
    \fi
    \if@ACM@countrypresent\else
        \ClassWarningNoLine{\@classname}{No country present for an affiliation}%
    \fi
}
\makeatother

\usepackage[english]{babel}
\usepackage{blindtext}
\usepackage{epsfig,endnotes}
\usepackage{subcaption}
\usepackage{graphicx}
\usepackage{amsmath}
\usepackage{float}
\usepackage[textfont=it]{caption}
\usepackage{wrapfig}
\usepackage{multirow, multicol, booktabs}
\usepackage[labelfont=sc]{caption}
\usepackage{graphbox}
\usepackage{lipsum}
\usepackage[linesnumbered,algoruled,boxed,lined]{algorithm2e}
\usepackage{algpseudocode}
\usepackage[export]{adjustbox}
\usepackage{bm}
\usepackage{booktabs}
\usepackage{titlesec}
\hypersetup{colorlinks=true, linkcolor=blue, urlcolor=blue}
\usepackage{dsfont}
\usepackage{adjustbox}
\usepackage{siunitx}
\usepackage{hyperref}
\usepackage{array, tabularx}
\usepackage{pifont}
\newcommand{\cmark}{\textcolor{green!60!black}{\ding{51}}}
\newcommand{\xmark}{\textcolor{red}{\ding{55}}}

\titlespacing*{\section}{4pt}{12pt}{12pt}
\titlespacing*{\subsection}{4pt}{12pt}{12pt}

\begin{document}

\newcommand{\name}{PeepLoc}
\newcommand{\para}[1]{\vspace{8pt}\noindent\textbf{#1}}

\newcommand{\dv}[1]{\textcolor{red}{DV: #1}}
\newcommand{\red}[1]{\textcolor{red}{#1}}

\renewcommand\footnotetextcopyrightpermission[1]{} 
\setcopyright{none}
\settopmatter{printacmref=false, printccs=false, printfolios=true}

\renewcommand{\shortauthors}{X.et al.}

\date{}

\title[Crowdsourcing Ubiquitous Indoor Localization with Non-Cooperative Wi-Fi Ranging]{Crowdsourcing Ubiquitous Indoor Localization with \\ Non-Cooperative Wi-Fi Ranging}

\author{
  Emerson Sie\textsuperscript{*}, 
  Enguang Fan\textsuperscript{*}, 
  Federico Cifuentes-Urtubey, 
  Deepak Vasisht
}
\affiliation{
  \institution{University of Illinois Urbana-Champaign}
}
\thanks{\textsuperscript{*}These authors contributed equally to this research.}

\begin{abstract}
Indoor localization opens the path to potentially transformative applications. Although many indoor localization methods have been proposed over the years, they remain too impractical for widespread deployment in the real world. In this paper, we introduce \name{}, a deployable and scalable Wi-Fi-based solution for indoor localization that relies only on existing devices and infrastructure. Specifically, \name{} works on any mobile device with an unmodified Wi-Fi transceiver and in any indoor environment with a sufficient number of Wi-Fi access points (APs) and pedestrian traffic. At the core of \name{} is (a) a mechanism which allows any Wi-Fi device to obtain non-cooperative time-of-flight (ToF) to any Wi-Fi AP and (b) a novel bootstrapping mechanism that relies on pedestrian dead reckoning (PDR) and crowdsourcing to opportunistically initialize pre-existing APs as anchor points within an environment. We implement \name{} using commodity hardware and evaluate it extensively across 4 campus buildings. We show \name{} leads to a mean and median positional error of 3.41 m and 3.06 m respectively, which is superior to existing state-of-the-art deployed indoor localization systems and is competitive with commodity GPS in outdoor environments.
\end{abstract}

\settopmatter{printacmref=false}
\setcopyright{none}
\renewcommand\footnotetextcopyrightpermission[1]{}
\pagestyle{plain}

\maketitle


\section{Introduction}
\label{sec:intro}

\begin{figure}
    \centering
    \includegraphics[width=.8\linewidth]{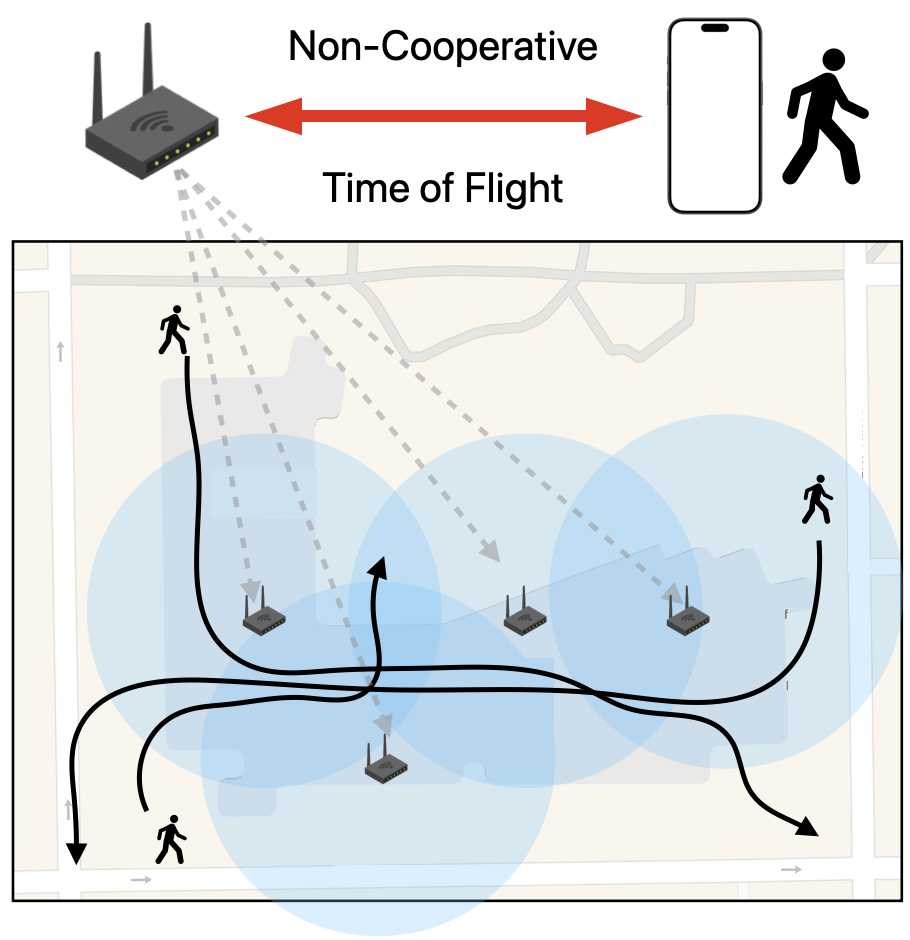}
    \caption{Non-cooperative time-of-flight measurements (i.e. one-way ranging) enables ranging between any Wi-Fi AP and client pair without specialized hardware requirements or modifications.}
    \label{fig:eyecatch}
    \vspace{-0.1in}
\end{figure}

Creating a wireless indoor localization system that can be ubiquitously deployed in the real world has been a long‐standing goal in the field. One highly viable path to such a system is through large-scale crowdsourcing using off-the-shelf smartphones. Most deployed crowdsourcing approaches today rely on opportunistically crowdsourced fingerprinting of Wi-Fi received signal strength (RSS) scans collected by users’ phones to build radio maps of indoor environments \cite{ni_experience_2022, apple_indoor_maps, rye_surveilling_2024}. Such systems suffer the classical disadvantages associated with RSS-based localization methods e.g. low accuracy and high sensitivity to noise and interference. Furthermore, they also suffer from the lack of scalability of fingerprinting-based methods (due to the human effort required).

Hence, recent efforts towards realizing ubiquitous Wi-Fi localization services have focused on Wi-Fi ranging for commodity devices e.g. 802.11mc Fine-Time Measurement (FTM) \cite{IEEE802.11mc2016}. Such ranging measurements are obtained by fine-grained time-of-flight (ToF) measurements of frames exchanged between a client device and responder (typically Wi-Fi AP). The basic principle of commodity Wi-Fi ranging is as follows. Consider a simple scenario where device A transmits a frame addressed to device B and records the round-trip time (RTT) upon receiving an ACK from device B. Then we expect \vspace{0.1in}

\begin{equation}\label{eq:tof-ranging}
    t_{\text{RTT}} = \frac{2d}{c} + t_{\text{proc}} \vspace{0.1in}
\end{equation}

\begin{figure*}[t]
    \centering
    \includegraphics[width=1.0\linewidth]{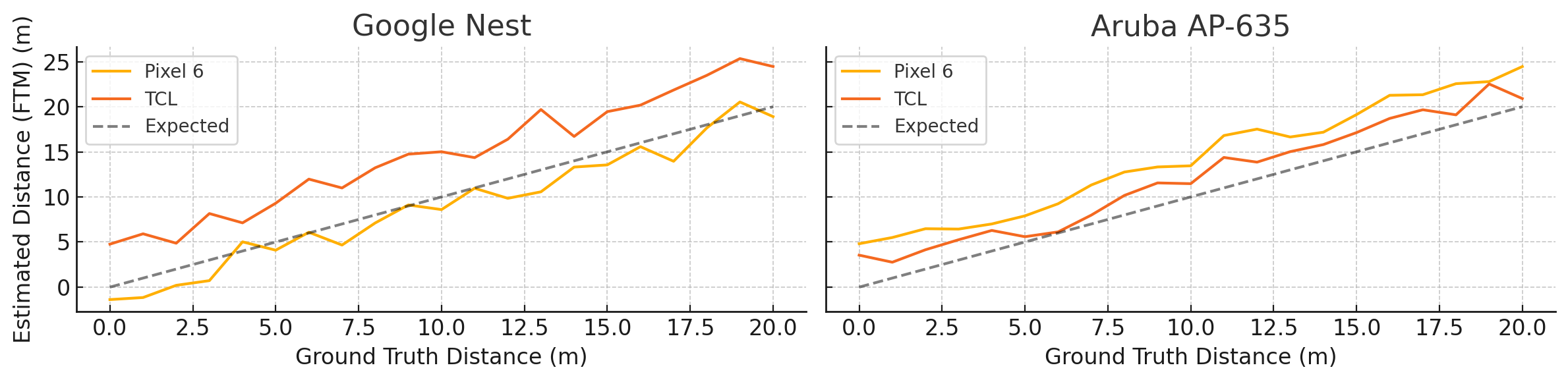}
    \caption{Effect of device heterogeneity on FTM ranging offsets. We experiment with two different client devices and two different APs. Changing either the client device or responding AP can change the ranging behavior.}
    \label{fig:device-heterogeneity}
\end{figure*}

where $t_{\text{RTT}}$ is the observed RTT at the initiating (client) device A, $d$ is the physical distance between the two devices, and $t_{\text{proc}}$ is the processing (offset) time at the responding device B. In general, $t_{\text{proc}}$ is unknown, but can be determined from implicit means (e.g. WiPeep \cite{abedi_non-cooperative_2022}) or explicit means (e.g. FTM protocol messages if supported).

Ranging measurements confer several advantages. First, ranging measurements promise finer-grained accuracy and better robustness to noise and interference than RSS-based methods. Second, ranging-based localization systems promise significantly better scalability than fingerprinting-based methods as only information about a sparse set of anchor points need to be stored for a particular environment. Yet, there are several key challenges preventing Wi-Fi ranging-based indoor positioning systems from widespread deployment.

\begin{itemize}

\item \textbf{Accurately Locating APs.} Having a large set of anchor points with known location is the foundation of any ranging-based localization system. Hence, the first challenge is in accurately geolocating a large number of Wi-Fi APs in a building to use as anchor points. Currently, this can be done through either (a) manual site survey, (b) RSS-based crowdsourcing methods \cite{ni_experience_2022,rye_surveilling_2024}, and (c) GPS-based systems such as Aruba OpenLocate \cite{troymart_its_2023}. Each of these methods have their respective disadvantages. First, site surveying lacks scalability as it requires dedicated human effort and is prone to human error. Second, RSSI-based systems inherit the disadvantages of RSSI-based ranging models. Finally, the OpenLocate model relies on APs having GPS receivers, which is uncommon. Moreover, it relies on APs being able to receive GPS readings indoors, which is not always possible. Furthermore, as APs are static when averaging GPS readings, they are prone to systematic multipath errors when self-localizing, further contributing error. In conclusion, there is a lack of a scalable and accurate AP geolocation method, which adversely affects Wi-Fi ranging-based localization systems as inaccuracies in AP locations will contribute systemic errors to client localization.

\item \textbf{Device Heterogeneity.} When conducting Wi-Fi ranging measurements between different devices, they exhibit different ranging offsets. This is true for both cooperative ranging (802.11mc FTM) and non-cooperative ranging (one-way). To see this for FTM, we conduct an experiment with two FTM-enabled clients (a Pixel 6 and TCL) and two FTM-enabled APs (Google Nest, Aruba AP-635). We show the results in Fig.~\ref{fig:device-heterogeneity}. As the figure suggests, different pairs of clients will feature different fixed range offsets even though the FTM protocol is designed to explicitly inform the client of an AP's offset. This behavior is consistent with the study by Agarwal et. al. \cite{aggarwal_is_2022}. Similarly, we note that in non-cooperative ranging, the measurements are affected by device-specific short interframe spacing (SIFS) values \cite{abedi_non-cooperative_2022}. To summarize, device heterogeneity presents a challenge to any Wi-Fi ranging based localization system as it presents offsets that much be calibrated for each client-responder pair.

\item \textbf{NLOS Ranging Errors.} Although the ranging model presented in Eq.~\ref{eq:tof-ranging} is correct in principle, we find that the observed time $t_{\text{RTT}}$ only matches this model in line-of-sight (LOS) or near LOS environments. In highly NLOS environments, $t_{\text{RTT}}$ is systematically higher than expected. This is due to additional processing overhead at both the client and responding device in lower RSSI environments (e.g. due to obstacle blockage). To confirm this, we conduct the following experiment. We take an AP and take ToF and RSSI measurements at varying distances from it in a LOS environment (Fig.~\ref{fig:tof-rssi-los}). Then, we place the AP behind two large walls and repeat the same experiment (Fig.~\ref{fig:tof-rssi-nlos}). We can see that in the NLOS case, the observed RSSI at each distance is lower than in the LOS case as expected. However, the observed ToF consistently overshoots the expected value for each distance. This overhead depends on the degree of signal blockers in an environment and presents an additional challenge for practical Wi-Fi ranging in buildings.

\begin{figure*}[t]
    \centering
    \begin{subfigure}[c]{0.49\linewidth}
        \centering
        \includegraphics[width=\linewidth]{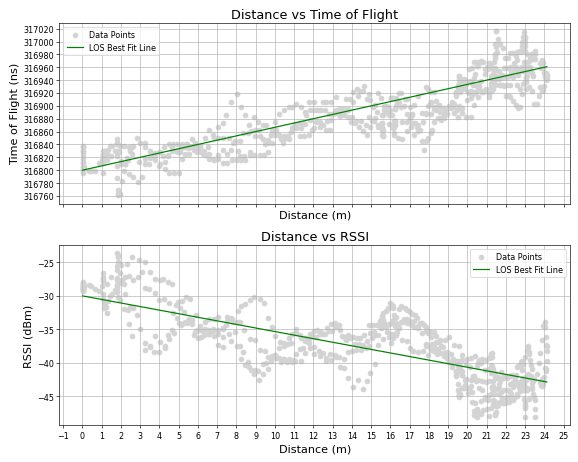}
        \caption{Distance vs ToF and RSSI in LOS (green).}
        \label{fig:tof-rssi-los}
    \end{subfigure}
    \hfill
    \begin{subfigure}[c]{0.49\linewidth}
        \centering
        \includegraphics[width=\linewidth]{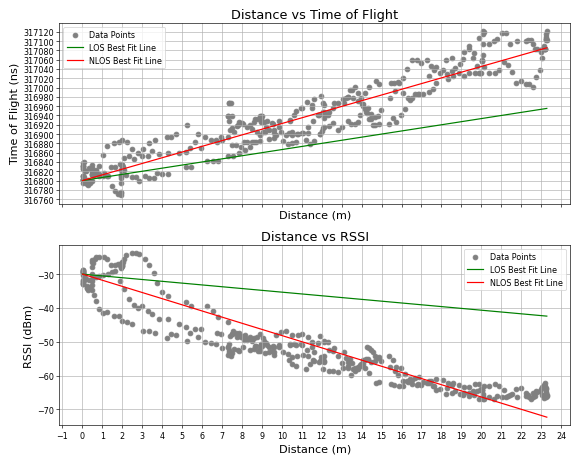}
        \caption{Distance vs ToF and RSSI in NLOS (red).}
        \label{fig:tof-rssi-nlos}
    \end{subfigure}
    \caption{Comparing one-way ranging in LOS and NLOS environments. At the same distance, a lower RSSI value leads to a higher observed ToF. }
    \label{fig:tof-rssi-los-nlos}
\end{figure*}

\end{itemize}

In this paper, we propose \name{}, which addresses these challenges through a combination of new techniques and system design. At the core of \name{} is (a) non-cooperative Wi-Fi ranging \cite{abedi_non-cooperative_2022}, which enables any Wi-Fi enabled device to conduct ranging measurements to any other Wi-Fi enabled device, and (b) accurate pedestrian dead-reckoning capabilities on modern smartphones, which we use to be able to harvest valuable information from large numbers of smartphone users regularly entering indoor environments. \name{} leverages these to bring the crowdsourcing framework to Wi-Fi ranging-based indoor positioning systems, enhancing accuracy, adaptability, and scalability. Moreover, while some prior indoor localization systems report high accuracy in controlled settings, they typically rely on specialized hardware, custom firmware, or labor-intensive site survey that severely limit their scalability and practicality in real-world environments. In contrast, \name{} is designed to operate entirely on commodity smartphones and unmodified Wi-Fi infrastructure. It does not require PHY-layer access, protocol support like 802.11mc/az, or prior knowledge of infrastructure layout. This “zero-assumption” design makes \name{} uniquely deployable at scale, turning ubiquitous Wi-Fi deployments and everyday pedestrian motion into a crowdsourced indoor localization system. In the following, we elaborate on three key aspects of \name{}.

\begin{itemize}

\item \textbf{Crowdsourcing APs with Non-Cooperative Ranging.} To overcome the scalability limitations of site surveying, the innaccuracies of RSS-based methods, and the hardware assumptions of GPS-based AP localization (e.g., OpenLocate), \name{} opportunistically collects non-cooperative Wi-Fi ranging measurements from smartphones on pedestrians as they enter environments in order to locate APs. We use non-cooperative Wi-Fi ranging over FTM as it works with any 802.11-compliant device-responder pair, meaning \name{} can potentially use any AP in the environment. The core idea is as follows. While a pedestrian is outdoors, \name{} can geolocate the smartphone using GPS, however upon entering a building, GPS is lost. \name{} detects the point at which GPS is lost and then switches to PDR (pedestrian dead reckoning) to continue getting accurate positioning (up to a certain point where intolerable drift will accumulate). While the trajectory of the smartphone is known, \name{} uses non-cooperative Wi-Fi ranging measurements to multilaterate and thus geolocate the positions of APs encountered. Over time, repeated visits by multiple users from diverse paths improve localization accuracy via spatial averaging. This approach avoids the need for dedicated setup efforts, leverages everyday motion for mapping, and naturally captures environmental diversity, improving AP localization in cluttered or dynamic environments.

\item \textbf{Infrastructure-Independent Operation.} As mentioned previously, crucial to the viability of Wi-Fi ranging is the accurate determination of $t_\text{proc}$ in Eq.~\ref{eq:tof-ranging}. Although there are means of explicitly determining $t_\text{proc}$ such as in FTM, in reality few devices support the protocol, and moreover among the limited devices that do, offset-related errors still exist. We conclude that some degree of offset determination will always exist between any distinct client-responder pair. Hence, we design \name{} to work independently of infrastructure capabilities. Unlike FTM-based ranging, we do not rely on network infrastructure support for determining ranging offsets. Instead, we move all ranging offset-related logic to run solely on the client/server side. This makes \name{} deployable with any commodity AP and device pair, seamlessly managing heterogeneity.

\item \textbf{NLOS Adaptation via Per‐AP Ranging Model.} With the locations and offsets of the APs known, the same model in Eq.~\ref{eq:tof-ranging} can be used to multilaterate the position of the client. However, this is more challenging than localizing the APs for the following reason. Smartphones on pedestrians can continuously aggregate ranging measurements to nearby APs while the pedestrian is in motion. With a large and diverse enough set of pedestrian trajectories (which is not difficult to obtain), we effectively have ranging measurements from a very high number (e.g., 1000s) of virtual anchor points, making the problem of multilaterating the APs highly over-determined. On the other hand, when localizing a client, we can only obtain ranging measurements from a much sparser set of fixed anchor points in the environment (e.g. less than 10 APs within range). This motivates us to improve the ranging model in Eq~\ref{eq:tof-ranging} as much as possible when localizing the client. As explained previously, the slope term $\frac{2}{c}$ in Eq.~\ref{eq:tof-ranging} underestimates the observed slope in NLOS environments, with the extent of the deviation depending on both the device and characteristics of its environment. Hence, \name{}'s ranging model incorporates a per-AP slope term during  trajectory fitting to capture environmental blockage and multipath characteristics, ensuring the ranging model remains robust even when \(t_{\mathrm{RTT}} > \frac{2d}{c} + t_{\mathrm{proc}}\).

\end{itemize}

\renewcommand{\tabularxcolumn}[1]{>{\centering\arraybackslash}m{#1}}

\definecolor{LightGreen}{rgb}{0.88,1,0.88} 

\begin{table*}[!t]
\vspace{-0.1in}
\centering
\caption{\label{tab:prior_work_comparison}Comparison of \name{} against prior work/existing systems.}
\begin{tabularx}{\linewidth}{|c|*{6}{X|}}
\hline
\textbf{System} &
\textbf{No AP Cooperation Needed} &
\textbf{No PHY Access Required} &
\textbf{No Prior AP Locations or Map Needed} &
\textbf{No Special Hardware Needed} &
\textbf{Data Collection Overhead} &
\textbf{Deployability} \\
\hline
RADAR/Horus/Zee \cite{bahl2000radar,youssef2005horus,chintalapudi_zee_2012} & \cmark & \cmark & \xmark & \cmark & High & Medium \\
ArrayTrack/SpotFi \cite{xiong_arraytrack_2013,kotaru2015spotfi} & \xmark & \xmark & \xmark & \xmark & Medium & Low \\
Chronos \cite{vasisht2016chronos} & \cmark & \xmark & \xmark & \xmark & Low & Low \\
Horn's System \cite{horn_indoor_2022} & \cmark & \cmark & \xmark & \cmark & Medium & Low \\
Wi-Fi RTT (802.11mc/az) \cite{IEEE802.11mc2016, IEEE802.11az2022, ibrahim2018rtt} & \xmark & \cmark & \cmark & \xmark & Low & Low \\
\rowcolor{LightGreen}
\textbf{\name{} (Ours)} & \textbf{\cmark} & \textbf{\cmark} & \textbf{\cmark} & \textbf{\cmark} & \textbf{Low} & \textbf{High} \\
\hline
\end{tabularx}
\end{table*}

We build a \name{} prototype using commodity off-the-shelf (COTS) hardware. To evaluate \name{}, we conduct extensive experiments in 4 different campus buildings and with devices from various different manufacturers. Our experiments show that, even with a small number of pedestrian trajectories, we are able to localize APs with a mean accuracy of 1.43 m and localize clients with a mean accuracy of 3.41 m.

In summary, our work makes the following contributions:
\begin{itemize}
    \item We propose a novel crowdsourcing method for fine-grained geolocation of Wi-Fi APs in indoor environments using non-cooperative Wi-Fi ToF measurements, GPS, and PDR on commodity smartphones. In this way, \name{} can bootstrap itself without dedicated human effort.
    \item We propose a new system design for Wi-Fi ranging-based indoor positioning systems, which involves moving offset determination logic out of the clients and network infrastructure onto a dedicated backend. This way, \name{} seamlessly manages device heterogeneity.
    \item We identify shortcomings of the naive Wi-Fi ranging model in NLOS indoor environments, namely slope deviation due to environmental factors. To address this, \name{} fits per-AP ranging models that improves accuracy and robustness in indoor environments.
    \item We contribute a toolkit for collecting non-cooperative Wi-Fi ranging measurements on commodity ESP32s. These are available at 
    \footnote{\url{https://github.com/ConnectedSystemsLab/esp32_wipeep_ros}}. We hope this drives further research into Wi-Fi ranging-based indoor positioning systems.
\end{itemize}

\section{Background \& Related Work}
\label{sec:background}

\subsection{Background}

\begin{figure*}[!t]
    \centering
    \begin{subfigure}[t]{0.3\linewidth}
        \centering
        \includegraphics[width=\linewidth]{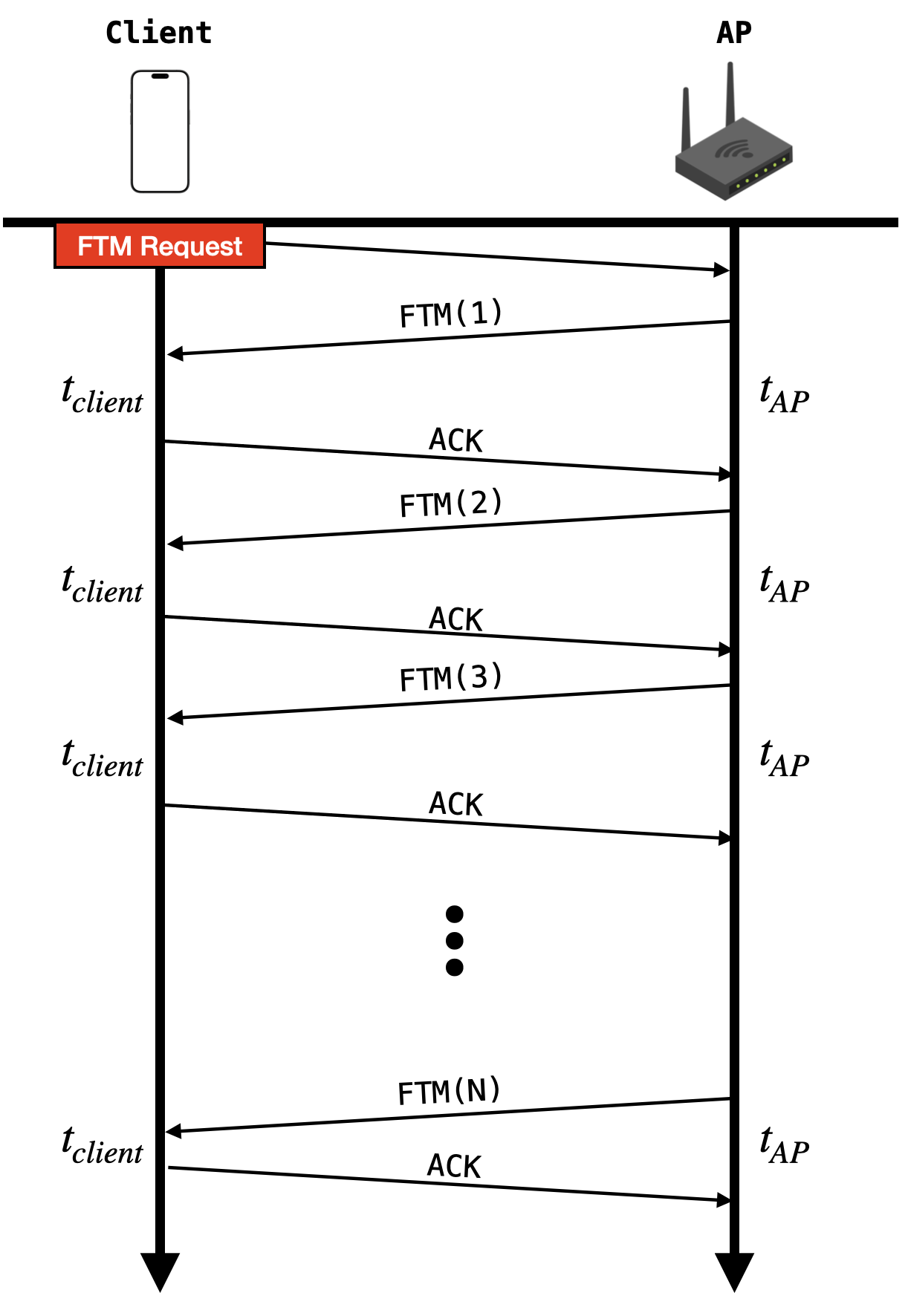}
        \caption{802.11mc (FTM) protocol.}
        \label{fig:2wr}
    \end{subfigure}
    \quad
    \begin{subfigure}[t]{0.3\linewidth}
        \centering
        \includegraphics[width=\linewidth]{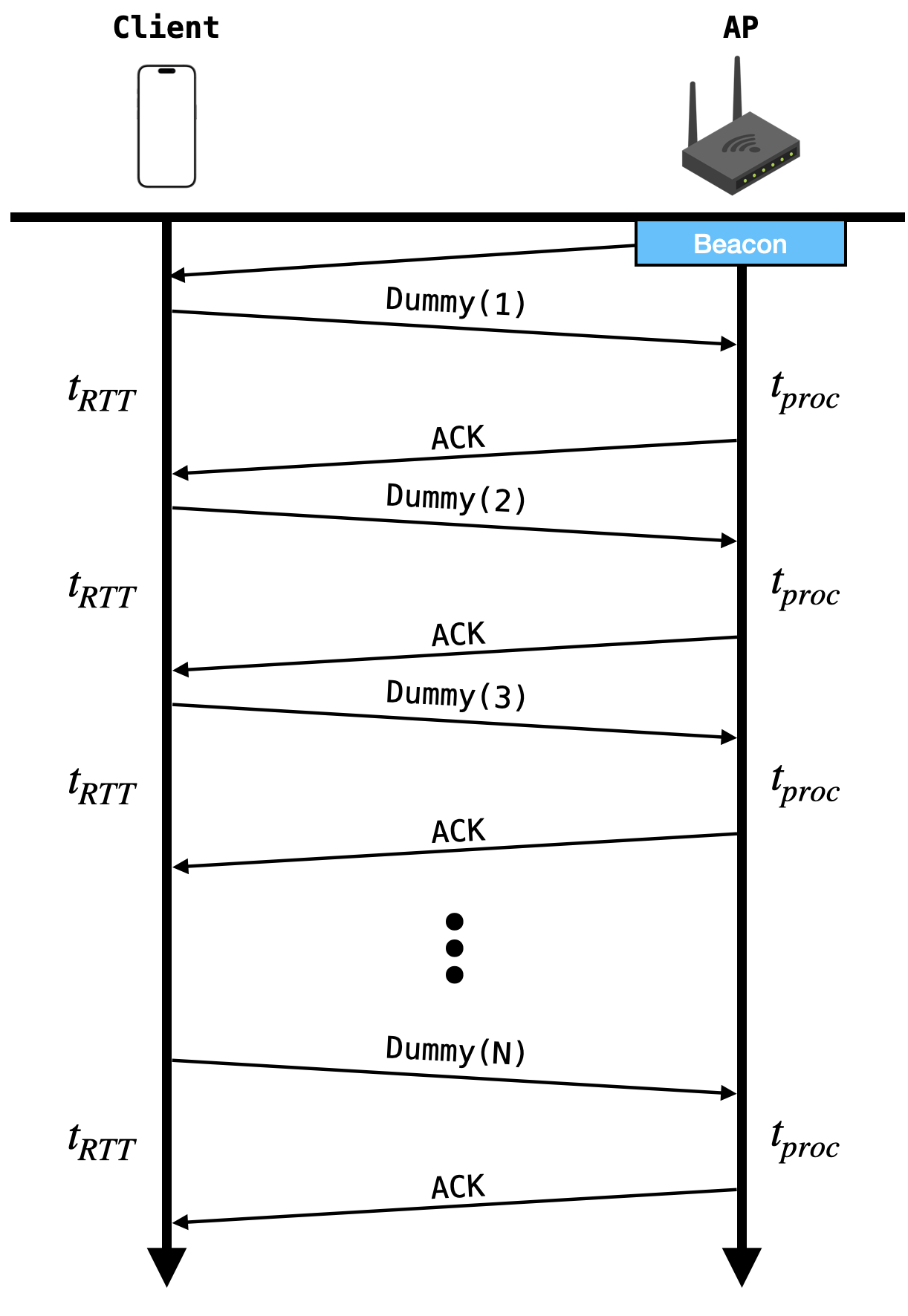}
        \caption{Non-Cooperative Wi-Fi ranging.}
        \label{fig:1wr}
    \end{subfigure}
    \caption{Comparison between 802.11mc (FTM) protocol and one-way Wi-Fi ranging. Note that the FTM protocol requires cooperation of both devices, whereas non-cooperative ranging works with any standards-compliant Wi-Fi device.}
    \label{fig:protocols}
\end{figure*}

\begin{figure*}[!t]
    \centering
    \begin{subfigure}[t]{0.51\linewidth}
        \centering
        \includegraphics[width=\linewidth]{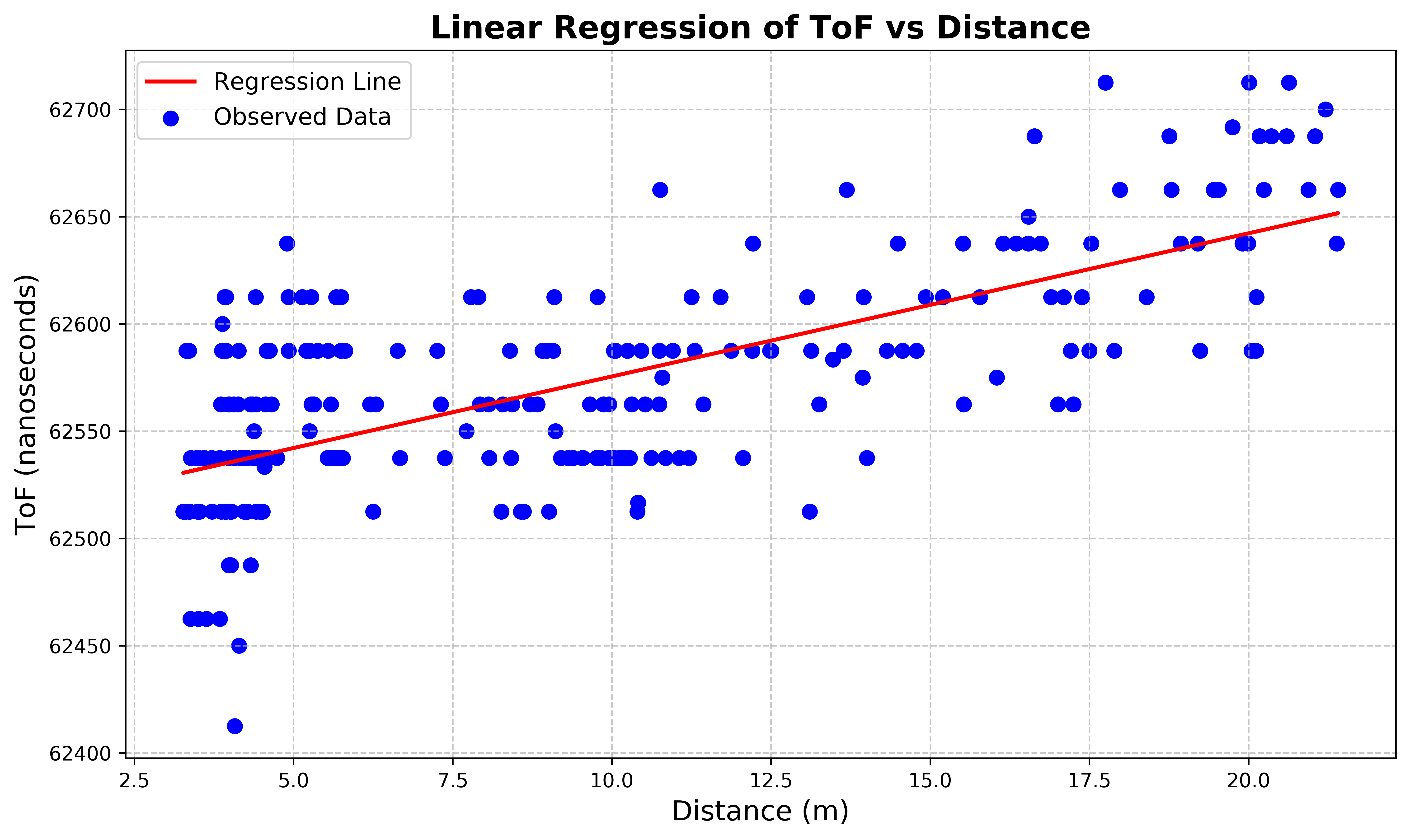}
        \caption{RTT vs distance in non-cooperative Wi-Fi ranging.}
        \label{fig:polite-wifi}
    \end{subfigure}
    \begin{subfigure}[t]{0.27\linewidth}
        \centering
        \includegraphics[width=\linewidth]{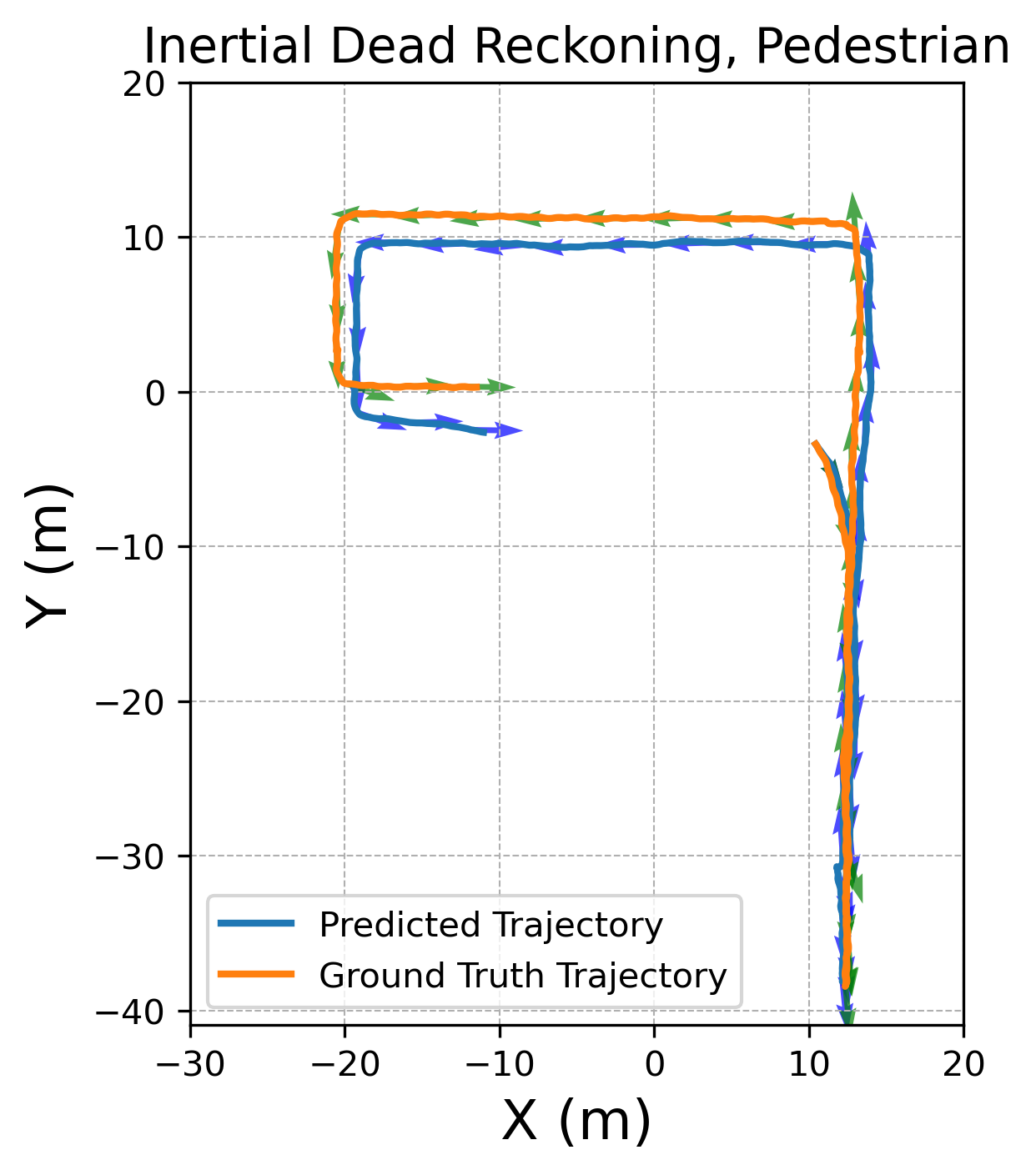}
        \caption{PDR on a $>100$ m long trajectory.}
        \label{fig:dead-reckoning}
    \end{subfigure}
    \caption{Sensing capabilities on modern smartphones.}
    \label{fig:sensing}
\end{figure*}

\para{Wi-Fi Ranging.}
Wi-Fi-based ranging primarily relies on the Fine Time Measurement (FTM) protocol, standardized in IEEE 802.11mc \cite{IEEE802.11mc2016} and extended in the newer IEEE 802.11az \cite{IEEE802.11az2022}. These protocols support both two-way and one-way RTT time-of-flight (ToF) estimation by exchanging timestamped packets between an initiating device (typically a client station like a smartphone) and a responding device (usually a Wi-Fi access point) (see Fig.~\ref{fig:2wr}). In the standard FTM procedure, the initiator sends a request package at timestamp $\textbf{t}_1$, the responder marks the reception timestamp and replies an ACK package containing precise reception timestamp $\textbf{t}_2$ and transmission timestamp $\textbf{t}_3$. Upon receiving the ACK package from the responsder at $\textbf{t}_4$, the precise round-trip time can be estimated by Eq. \ref{eq:two-sided-rtt}. By analyzing the round-trip time (RTT) across these frames and acknowledgments (ACKs), the initiator can estimate ToF and thus compute distance.
\begin{equation} \label{eq:two-sided-rtt}
\text{ToF}_{\text{2-sided}} = {(t_4 - t_1) - (t_3 - t_2)}
\end{equation}

While FTM-based ranging offers high accuracy, its adoption remains limited due to its dependence on hardware and firmware support at both ends of the link. Only a subset of commercial devices—primarily high-end smartphones (e.g., Google Pixel, Samsung Galaxy S series) and APs from vendors like Aruba—implement the required standards. As a result, widespread deployment is constrained by specialized hardware requirements. For responder like legacy Wi-Fi access point which lacks the precise timing capability, the FTM protocol degrades to one-way RTT, where responder doesn't provide the reception and transmission timestamps. In such cases, the initiator can only approximate the round-trip time as Eq. \ref{eq:one-sided-rtt} unless the AP’s turn-around offset can be estimated. 

\begin{equation} \label{eq:one-sided-rtt}
\text{ToF}_{\text{1-sided}} = {(t_4 - t_1)}
\end{equation}

In contrast, non-cooperative ranging presents a lightweight alternative that operates on any standards-compliant Wi-Fi device without relying on 802.11mc/az (see Fig.~\ref{fig:1wr}). In this mode, device A transmits a dummy data packet (Null Data Packet) to device B and measures the RTT based on the ACK response from device B. However, since the receiver’s internal processing delay is not reported, estimating ToF becomes challenging. Systems like WiPeep \cite{abedi_non-cooperative_2022} address this by leveraging spatial diversity and passive RTT measurements from multiple vantage points to infer hidden delays and accurately estimate distance—enabling ranging without protocol support or device cooperation.

\para{Pedestrian Dead Reckoning.} Pedestrian Dead Reckoning (PDR) is a navigation technique that estimates a pedestrian’s position by utilizing data from pedestrian-mounted Inertial Measurement Units (IMUs) to track steps and heading changes. Traditional PDR methods employ analytical models and filtering techniques, such as Zero Velocity Updates (ZUPTs) and Kalman filtering, to integrate sensor measurements and compute position and orientation. However, these methods often suffer from cumulative errors due to sensor drift, particularly when the IMUs are not mounted on the foot, where motion patterns are more predictable 

Recent advancements have introduced neural inertial methods that leverage deep learning models to learn complex motion patterns from inertial data, providing more robust and accurate position estimations. Notable examples include IONet \cite{chen_ionet_2018}, RoNIN \cite{herath_ronin_2020}, and TLIO \cite{liu_tlio_2020}. IONet utilizes Long Short-Term Memory (LSTM) networks to regress velocity and heading changes, mitigating drift over time. RoNIN employs deep neural networks to extract high-level motion features, enhancing positioning accuracy in diverse and dynamic environments. TLIO combines deep learning with an Extended Kalman Filter (EKF) framework, fusing learned displacement estimates and uncertainties to solve for pose, velocity, and sensor biases. These methods demonstrate that pedestrian motion data is amenable to learning-based approaches, leading to state-of-the-art PDR results. An example of such performance is illustrated in Fig.~\ref{fig:dead-reckoning}. For a comprehensive overview of progress in this area, we refer the reader to the survey by Chen et al. \cite{chen_deep_2024}.

\subsection{Related Work}
\label{sec:related}

\para{Localization with Wi-Fi Fingerprinting.} Fingerprinting remains one of the most widely adopted techniques for indoor localization. It leverages ambient signal measurements from various sensors (e.g. Wi-Fi, BLE, magnetometer) to create a map of the environment. The seminal RADAR system \cite{bahl2000radar} demonstrated that Wi-Fi RSS fingerprints could localize users with median errors of 2–3 meters in an office environment without additional hardware infrastructure. Subsequent work has improved on this framework using better statistical models e.g. Horus \cite{youssef2005horus}, including more sensors \cite{wang_no_2012}, incorporating crowdsourcing e.g. Zee \cite{chintalapudi_zee_2012}, and incorporating deep learning \cite{nowicki2017wifi}. Fingerprinting has resulted in many commercial-grade crowdsourced systems (e.g. Apple's indoor maps program \cite{apple_indoor_maps} and Tencent's indoor localization system \cite{ni_experience_2022}). Despite their popularity, fingerprinting systems have well known challenges. First, they require costly and time-consuming data collection. Second, they require that floorplans or maps of the environment are known beforehand. Finally, their performance degrades in dynamic environments.

\para{Localization With Wi-Fi PHY Layer Information.} Another well-researched class of localization systems infer location from observing the physical channel at either the client device or the anchor points (Wi-Fi APs). Representative examples include ArrayTrack \cite{xiong_arraytrack_2013}, SpotFi \cite{kotaru2015spotfi}, and and Chronos \cite{vasisht2016chronos}. ArrayTrack \cite{xiong_arraytrack_2013} uses PHY information from antenna arrays at MIMO APs to measure AoA towards clients and triangulate their location. SpotFi \cite{kotaru2015spotfi} extends this framework by adding MUSIC super-resolution and clustering to find the LOS path. Breaking away from the AoA framework, Chronos \cite{vasisht2016chronos} instead measures channel estimates across multiple frequency bands to estimate ToF between a device and AP pair. Although these methods lead to competitive localization accuracy, they suffer from several drawbacks. First, they require access to low-level PHY data and customized firmware. Second, they require known AP locations and orientations in the case of AoA-based systems. These factors limit their practical deployment.

\para{Localization with Cooperative Ranging.} To overcome the disadvantages of fingerprinting and PHY-reliant methods, time-of-flight (ToF) based methods have been proposed beginning with the first two-way Wi-Fi ranging standard IEEE 802.11mc (and later 802.11az). These protocols require implementation across both Wi-Fi APs and clients to enable out-of-the-box estimation of round-trip propagation delay. Ibrahim et al. evaluated the feasibility of Wi-Fi RTT on commercial smartphones and achieve sub-2 meter accuracy in line-of-sight indoor settings \cite{ibrahim2018rtt}. Although two-way Wi-Fi ranging has high accuracy, it is supported only by a few devices and AP manufacturers on the market (e.g. Aruba Open Locate \cite{troymart_its_2023}), limiting its widespread deployment.

\para{Localization with Non-Cooperative Wi-Fi Ranging.} Unlike two-way (cooperative) ranging methods, non-cooperative ranging methods do not rely on cooperation at the AP. These include our system, Wi-Peep \cite{abedi_non-cooperative_2022}, and Horn's system \cite{horn_indoor_2022}. Wi-Peep tackles the problem of localizing devices inside an indoor environment using a GPS-equipped drone flying around the building. Since the drone functions as the anchor point for localization in this case, it is not suitable as a permanently deployed localization system. The system proposed by Horn \cite{horn_indoor_2022} uses non-cooperative Wi-Fi ToF as an indoor localization system, but assumes the locations of the APs within the environment are known beforehand. By contrast, our system is capable of automatically locating where the APs are inside an environment through crowdsourcing.

\para{\name{} in Context.}
As summarized in Table~\ref{tab:prior_work_comparison}, \name{} stands out from prior work by eliminating common deployment barriers across both cooperative and non-cooperative localization systems. Unlike fingerprinting approaches (RADAR\cite{bahl2000radar}, Horus\cite{youssef2005horus}, Zee \cite{chintalapudi_zee_2012}) that require labor-intensive data collection and known maps, or PHY-dependent systems (ArrayTrack \cite{xiong_arraytrack_2013}, SpotFi \cite{kotaru2015spotfi}, Chronos \cite{vasisht2016chronos}) that rely on firmware or hardware modifications, \name{} operates without any specialized support from the access points, physical layer access, or knowledge of infrastructure layout. Moreover, unlike prior one-way ranging systems such as WiPeep \cite{abedi_non-cooperative_2022} or Horn’s system \cite{horn_indoor_2022}, \name{} does not rely on external infrastructure (e.g., drones or GPS) or assume pre-calibrated AP locations. Instead, it introduces a novel system design that is fully software-defined and self-contained: all ranging and inference are performed using standard Wi-Fi frames and a commodity smartphone alone. This softwarized approach, requiring no firmware changes or hardware capabilities beyond what is already present in 802.11-compliant devices, enables true plug-and-play deployability. In doing so, \name{} advances the state of the art by bridging the gap between deployability, compatibility, and accurate ranging—all within an unmodified wireless stack.

\section{Design}
\label{sec:design_choices}

\begin{figure*}[!t]
    \centering
    \includegraphics[width=1.0\linewidth]{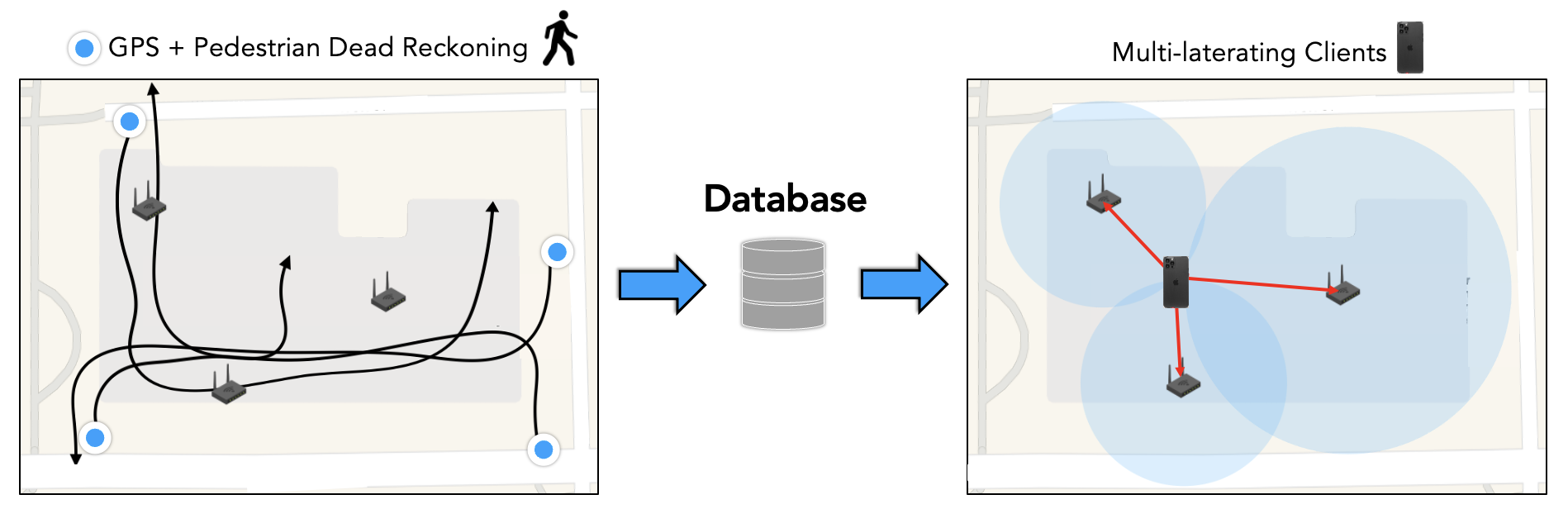}
    \caption{Overview of \name{}. In the first phase (left), \name{} uses GPS, PDR, and one-way ranging to locate all Wi-Fi APs in a building. In the second phase (right), a client device uses  one-way ranging and known AP locations to self-localize.} \vspace{0.1in}
    \label{fig:system-overview}
\end{figure*}

\subsection{Goals}

\para{Design Goals.} We aim to create a wireless localization system that can work in virtually any indoor environment. To accomplish this, our system leverages three observations. First, there are large numbers of Wi-Fi APs in virtually all modern indoor environments (e.g. residential areas, corporate offices, public spaces like malls). These provide the ideal anchor points for ubiquitous localization. Second, there are approximately 4.88 billion smartphone users worldwide. Many of such devices have sophisticated sensing capabilities, such as (a) always-on motion sensing constantly running in the background (e.g. step counting for health tracking) and (b) on-board GPS, offering  meter-level outdoor geolocation on consumer-grade devices. Finally, users carry their phones in and out of buildings each day as they live their lives, offering abundant and exhaustive information about said environments.  

Accordingly, our goal is to opportunistically use crowdsourced data from pedestrian-mounted smartphones to initialize the localization system. First, our system must geo-locate the APs from pedestrian trajectories, initializing them as anchor points associated with a particular building and storing them in a database. Once the system has been initialized, a client entering a building can retrieve known anchor points from the database and listen to surrounding APs to determine its location. 

\subsection{Design Choices}

In the following, we elaborate on the rationale underlying the design choices of \name{}:

\para{Advantages of Wi-Fi Ranging.} For both AP and client localization, our system uses Wi-Fi ranging measurements over alternatives such as RSS and AoA for the following reasons. First, RSS ranging measurements are inaccurate due to coarse spatial resolution and sensitivity to noise/interference/Wi-Fi chip. Second, AoA-based triangulation would require multiple antennas with known geometries at the client (necessitating hardware overhead) and would rely on accurate device orientation to pinpoint the location of APs (limiting overall accuracy). By leveraging Wi-Fi ToF measurements, we can sidestep the shortcomings of RSS-based methods and the excessive requirements of AoA-based methods.

\para{Use of Pedestrian Trajectories.} Pedestrians are ideal for data collection over alternative automated solutions such as robots for site-surveying \cite{ayyalasomayajula_locap_nodate,arun_viwid_2022} for the following reasons. First, pedestrian data is much more scalable due to the disproportionately larger number of pedestrians roaming indoor environments in the world compared to robots. Moreover, as indoor environments are inherently human-centric, pedestrians can access more of them than robots can (e.g. by opening doors and traversing stairs with ease). Second, the rhythmic leg motion of pedestrians makes it the ideal platform for dead reckoning, unlike robots which typically require a combination of visual (e.g. cameras) and non-visual (e.g. IMUs, wheel encoders) sensors to perform trajectory tracking.

\para{Feasibility of Crowdsourcing.} Is large scale crowdsourcing of data from smartphones feasible? Here, we note that Wi-Fi crowdsourcing is already being done at the OS level on both major mobile platforms (i.e. Location Accuracy with Android, CoreLocation with iOS) \cite{rye_surveilling_2024}. In both systems, information about Wi-Fi APs are stored in web database. Hence there is already a pre-existing framework for our system to operate.

\para{Reducing Infrastructure Dependence.} Current state-of-the-art Wi-Fi ranging systems rely on infrastructure capabilities to deal with ranging offsets (e.g. FTM). However, as shown in Fig.~\ref{fig:device-heterogeneity}, offsets still remain regardless of FTM, and moreover offsets vary with differing client-responder pairs due to differing vendor implementations of the FTM protocol. We argue that pushing software updates to network infrastructure to correct the issue is too cumbersome. Instead, we move offset determination logic out of the infrastructure and into a separate backend i.e. the same one used to store information about APs \cite{rye_surveilling_2024}. We then use crowdsourcing to determine offsets for client-responder pairs in an online manner --- in this way, \name{} seamlessly manages device heterogeneity.

\para{Privacy Concerns.} Does our system raise privacy concerns? Although some spaces are public (i.e. malls, airports, subway stations), some might be private (i.e. office buildings). This raises a concern whether our system will use certain APs without consent. Here, we note that our system does not need to store AP information in a publicly accessible database where it is shared between mutual strangers (e.g. \cite{rye_surveilling_2024}). Instead, it is possible to deploy our system in a client-specific configuration where all APs are stored privately on-device and thus tailored to the specific buildings the a user visits.

\section{System}
\label{sec:design}

\subsection{System Overview}

We describe a high-level overview of \name{}. As shown in Fig.~\ref{fig:system-overview}, \name{} proceeds in two phases:

\begin{itemize}
\item \textbf{AP Discovery \& Localization.} In this phase, \name{} discovers and locates APs in an indoor environment as anchor points. It does this by opportunistically collecting data from smartphones on pedestrians as they enter environments. While a pedestrian is outdoors, we can easily geo-locate the smartphone using GPS. When the pedestrian enters a building, GPS is lost however we can still continue to get accurate positioning using inertial dead reckoning (up to a certain point where intolerable drift will accumulate). While the position of the smartphone is known, \name{} uses non-cooperative time-of-flight ranging to solve for the parameters of the APs encountered (location, offset, slope). The parameters of the APs are then recorded in a database.

\item \textbf{Client Localization.} In this phase, \name{} operates to localize clients within an indoor environment. When a client enters a particular building (as detected by GPS), it can retrieve the stored parameters of the APs inside the building (i.e. BSSIDs, locations, ranging model). It can then use its current ToF readings to nearby APs in order to multilaterate its own position within the building. The database used to store information about the APs can be shared (i.e. accessed over the Internet via cellular connection and shared with other clients) or private to the user (in which case, the database is on-device and used solely by the client).
\end{itemize}

\subsection{AP Discovery \& Localization}
\label{sec:dila-discovery}

In this phase, the system receives as input a collection of $n$ samples obtained from pedestrian trajectories. Each sample is represented as a tuple $(x_i, y_i, t_i, r_i, b_i)$, where $(x_i, y_i)$ denotes the estimated client location in caretsian Universal Transverse Mercator (UTM) coordinates, $t_i$ is the observed round-trip time (RTT), $r_i$ is the received signal strength indicator (RSSI), and $b_i$ is the BSSID of the access point (AP) involved in the measurement. When a GPS fix is available, the client's trajectory is estimated using GPS-IMU fusion. However, when the confidence of the GPS fix falls below a threshold, we assume the client has entered an indoor environment and switch to IMU-based PDR.

Then, our goal is to recover a set of AP parameters, denoted as $(x_j, y_j, c_j, b_j)_{j=1}^m$, where $(x_j, y_j)$ represents the physical location of AP $j$, $c_j$ is its processing delay, and $b_j$ is the associated BSSID.

For each AP $j$, we consider all measurements where $b_i = b_j$. Recalling the naive ranging model in Sec.~\ref{sec:intro}, we assume the RTT obeys: \vspace{0.1in}

\begin{equation}
    t_i = \frac{2d_i}{c} + c_j + \epsilon_i, \quad \epsilon_i \sim \mathcal{N}(0, \sigma^2), \vspace{0.1in}
    \label{eq:wipeep-model}
\end{equation}

where $d_i = \| (x_i, y_i) - (x_j, y_j) \|$ is the Euclidean distance between the client and AP, $c$ is the speed of light, and $\epsilon_i$ models Gaussian noise.

We use this basic model to infer key parameters of the APs. Specifically, given a candidate AP location and delay $(x, y, c)$, we define the maximum likelihood loss: \vspace{0.1in}

\begin{equation}
    \ell(x, y, c) = \sum_{i=1}^n \left\| t_i - \left( \frac{2}{c} \| (x_i, y_i) - (x, y) \| + c \right) \right\|^2. \vspace{0.1in}
    \label{eq:wipeep-loss}
\end{equation}

To localize AP $j$, we solve the following optimization problem: \vspace{0.1in}

\begin{equation}
    \underset{(x, y, c)}{\arg\min} \sum_{i=1}^{n} \left\| t_i - \left( \frac{2}{c} \| (x_i, y_i) - (x, y) \| + c \right) \right\|^2 \\ \cdot \mathds{1}\{b_i = b_j\}. \vspace{0.1in}
    \label{eq:wipeep-opt}
\end{equation}

where the search space for $(x, y)$ is constrained to the bounds of the building as estimated from satellite imagery, and $c$ is restricted to $[8, 12]~\mu$s in accordance with the 802.11 standard. In practice, we find this optimization problem well-behaved when averaging readings over 100-1000s of spatially-diverse sampling points, irrespective of the the inaccuracies of the ranging model (for example, due to NLOS environments). We repeat this for each unique BSSID observed in the dataset. The resulting AP parameters are recorded in a database for use during client localization.

\subsection{Client Localization}
\label{sec:dila-localization}

After AP discovery, the system proceeds to estimate the position of a mobile client. We assume the client has downloaded and cached metadata for nearby APs prior to entry into the building. Each AP $j$ is associated with parameters $(x_j, y_j, c_j,)$.

The key challenge in this stage is that unlike multi-laterating the APs, we often deal with only a sparse set of ranging measurements when attempting to localize a client (e.g. we can only detect 4 nearby APs). In such a scenario, the NLOS model errors of each individual ranging measurement (i.e. Eq.~\eqref{eq:wipeep-model}) have a far greater effect on the outcome. To amend this, we propose a NLOS-aware ranging model adapted for NLOS indoor environments as follows: \vspace{0.1in}

\begin{equation}
    t_{\text{RTT}} = \alpha \cdot d + t_{\text{proc}}, \vspace{0.1in}
    \label{eq:dila-model}
\end{equation}

where $d$ is the true distance between client and AP, $t_{\text{proc}}$ is the AP's processing delay, and $\alpha$ is a learned scale factor that models the RTT gradient under NLOS conditions. This parameter $\alpha_j$ is fit per AP using a similar process to Sec.~\ref{sec:dila-discovery} once the AP's other parameters (i.e. $(x_j, y_j, c_j)$) are known and fixed.

At runtime, the client collects a new set of $n$ measurements $\{(t_i, r_i, b_i)\}_{i=1}^n$, where each $t_i$ and $r_i$ correspond to RTT and RSSI from an AP with BSSID $b_i$. The client position $(x, y)$ is then estimated by minimizing the squared error between observed and predicted RTTs: \vspace{0.1in}

\begin{equation}
    (\hat{x}, \hat{y}) = \underset{(x, y)}{\arg\min} \sum_{i=1}^{n} \left\| t_i - \left( \alpha_{b_i} \cdot \| (x, y) - (x_{b_i}, y_{b_i}) \| + c_{b_i} \right) \right\|^2. \vspace{0.1in}
    \label{eq:dila-opt}
\end{equation}

This optimization yields the most likely client position based on available APs, their locations, and learned NLOS-adjusted ranging parameters.
\section{Implementation}
\label{sec:implementation}

\begin{table*}[ht]
\centering
\caption{Building floorplan dimensions and data collection summary}
\label{tab:floorplans}
\begin{tabular}{
  l
  S[table-format=3.1]
  S[table-format=3.1]
  S[table-format=4.0]
  S[table-format=4.0]
  S[table-format=2.0]
  S[table-format=2.0]
}
\toprule
\textbf{Floorplan} &
\textbf{Length (m)} &
\textbf{Width (m)} &
\textbf{\# Trajectories} &
\textbf{\# Test Points} &
\textbf{\# Netgear APs} &
\textbf{\# ASUS APs} \\
\midrule
A  & 94.17 & 37.40 & 10 & 11 & 5 & 4 \\
B  & 37.69 & 74.78 & 9 & 10 & 8 & 0 \\
C  & 26.12 & 50.49 & 9 & 6 & 8 & 0 \\
D1 & 30.24 & 78.45 & 11 & 10 & 9 & 0 \\
D2 & 30.24 & 78.45 & 14 & 15 & 4 & 3 \\
\bottomrule
\end{tabular}
\end{table*}

\begin{figure}[t]
    \centering
    \begin{subfigure}[b]{0.45\linewidth}
        \centering
        \includegraphics[width=\linewidth]{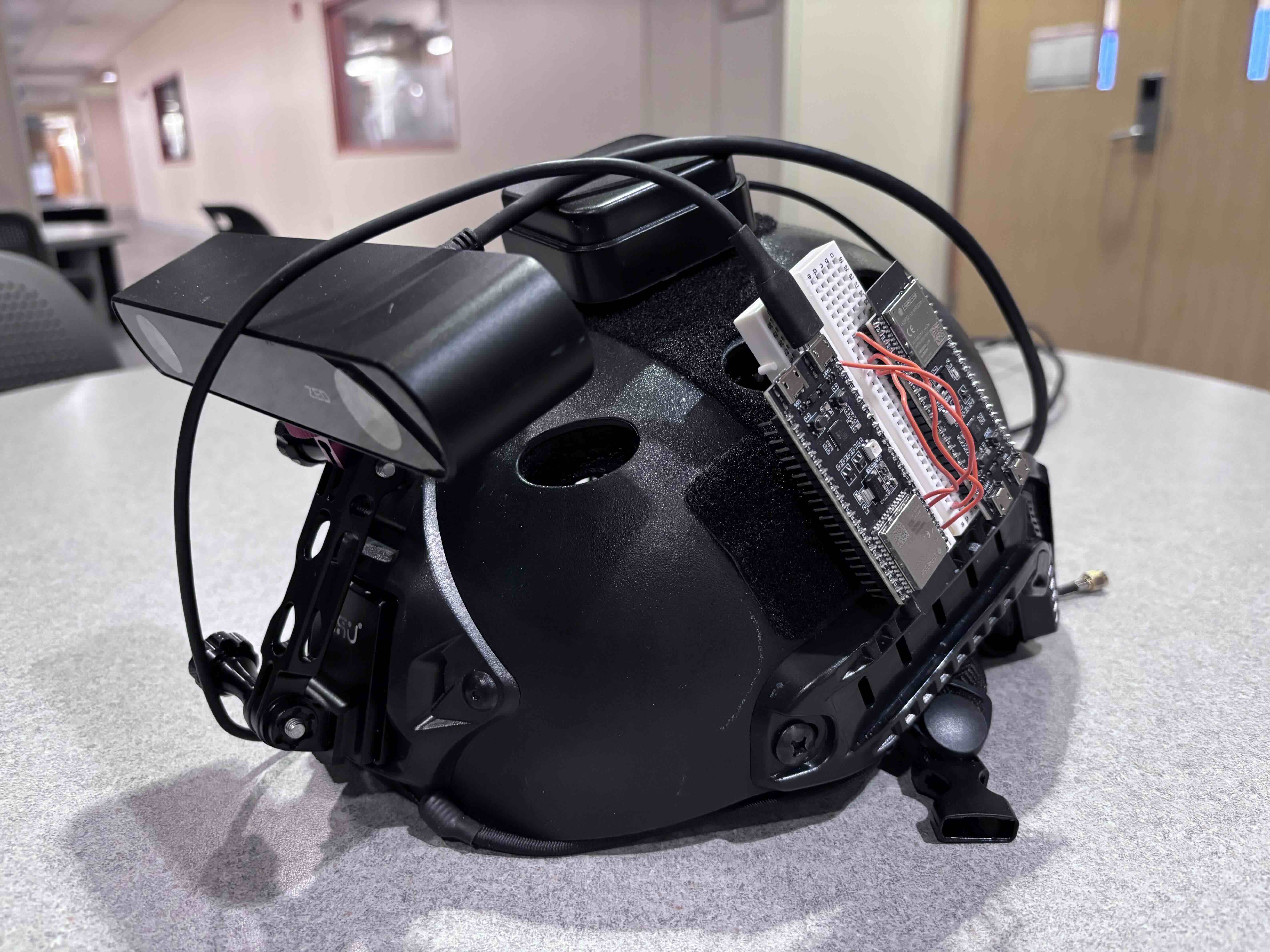}
        \caption{Sensors on helmet.}
        \label{fig:hardware}
    \end{subfigure}
    \hspace{0.05\linewidth}
    \begin{subfigure}[b]{0.45\linewidth}
        \centering
        \includegraphics[width=\linewidth]{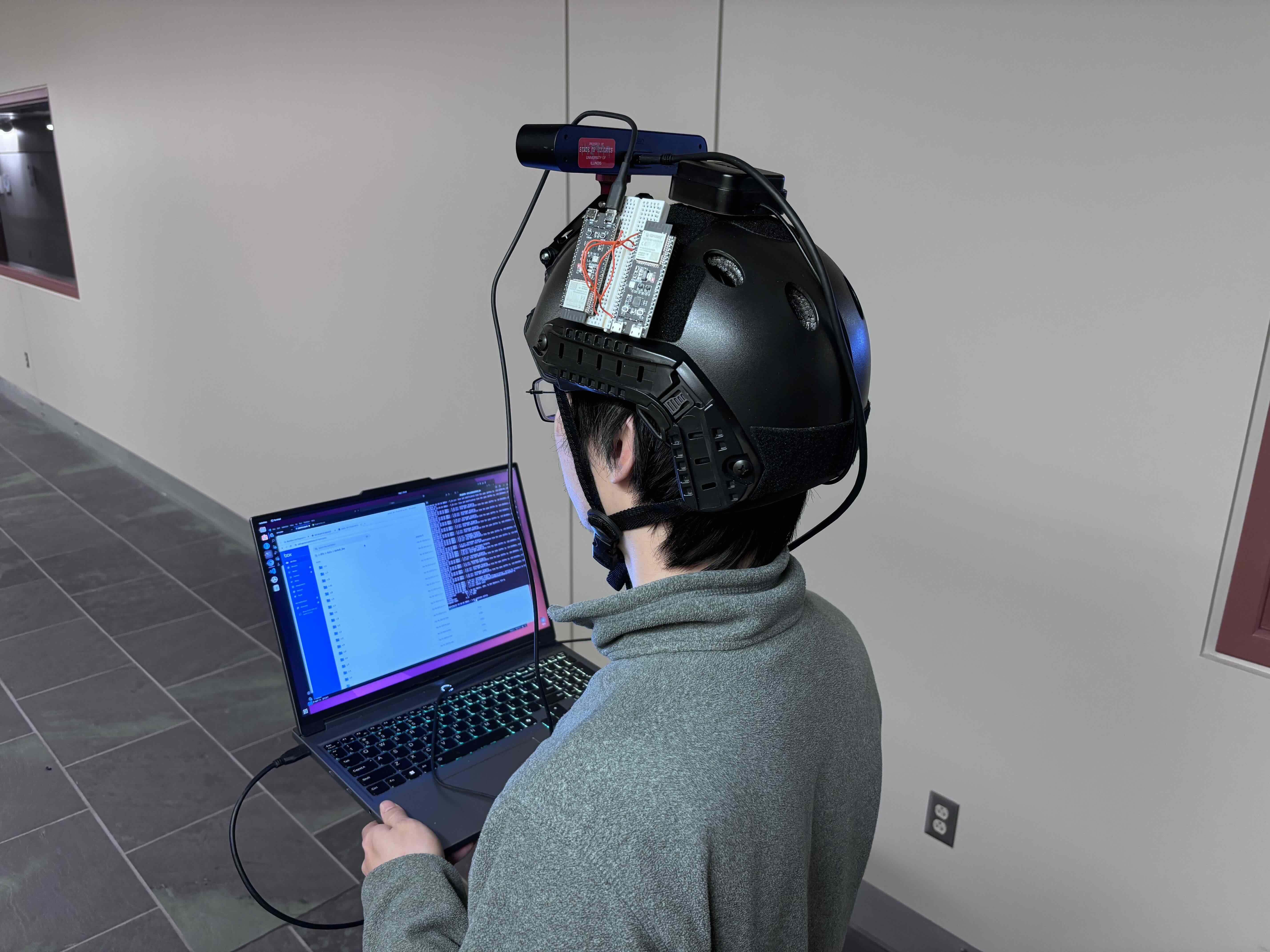}
        \caption{Helmet on pedestrian.}
        \label{fig:hardware-worn}
    \end{subfigure}
    \caption{Hardware. We attach a Zed 2i stereo camera, paired ESP32s, and ZED-F9P GPS to a helmet (which is worn when collecting data). This setup mimics the set of sensors found on commodity smartphones.}
    \label{fig:hardware-implementation}
\end{figure}

\para{Sensors.} On the client side, \name{} requires a standard commodity-grade Wi-Fi transceiver, IMU, and GPS receiver. These components can be found within virtually any modern smartphone. For practical reasons, we mimic a set of smartphone sensors using the discrete setup shown in Fig.~\ref{fig:hardware}. Our setup consists of a Zed 2i stereo camera (with built-in IMU) \cite{stereolabs_zed_2025}, a ZED-F9P GPS receiver \cite{ublox_zed_f9p_2023}, and two paired ESP32-S3s \cite{espressif_esp32_s3_2025} to measure time-of-flight and RSSI with nearby Wi-Fi APs. We use a SLAM algorithm on the Zed camera to obtain the positional ground truth of the client as it moves inside a building --- information from the camera is otherwise unused by \name{}. 

\para{Firmware.} We use two ESP32-S3 in tandem to implement the protocol shown in Fig.~\ref{fig:1wr}. The first ESP32 which behaves as the "injector" and the other ESP32 behaves as a "sniffer". Both are always set to be on the same 20 MHz band in the 2.4 GHz spectrum. The injector periodically sends probe requests to nearby APs while listening to and recording AP beacon frames. At regular intervals, the injector sends a sequence of dummy frames to each AP detected within the said interval (from beacon frames), ignoring beacon frames whose RSSI is below a certain threshold (we set a cutoff at -60 RSSI). The sniffer listens to these outgoing dummy frames and corresponding ACKs from the APs, recording the time difference between them (as well as the RSSI of the ACKs). To record the time difference between frame transmission and ACK reception, the sniffer counts the CPU clock cycles between the two events. In our case, we use an ESP32 with a clock frequency of 240 MHz --- this yields a time resolution of approximately $4.16$ ns, or $0.625$ m in distance terms. At the end of each interval, the sniffer reports these values to the host before repeating the process. This process aims to ensure that we have roughly the same number of readings for each AP in range across any given time interval. In terms of throughput, our setup can average up to 30 one-way ranging measurements per second to every AP within range.

\para{Software.} We run all code and algorithms on a laptop attached to the setup shown in Fig.~\ref{fig:hardware}. The laptop is intended to mimic the general compute capabilities of a smartphone attached to the sensors. To determine pedestrian trajectories outdoors, we use a GPS-IMU fusion algorithm \cite{stereolabs_global_2025}. We detect outdoor-indoor transitions by monitoring the confidence of the GPS fix as reported by the ZED-F9P and assigning a cutoff. Then, to determine dead-reckoning trajectories from pure IMU data, we use RNIN-VIO \cite{chen_rnin-vio_2021}. RNIN-VIO is a lightweight deep learning model for inertial PDR that works on embedded devices. We pre-train the model on the RNIN-VIO dataset and fine-tune the model on our dataset in order to take into account hardware differences.

\section{Evaluation}
\label{sec:evaluation}

\subsection{Methodology}

We evaluate \name{} across 5 different floorplan configurations spanning 4 different campus buildings (A,B,C,D), summarize in Table.~\ref{tab:floorplans}. Each floorplan configuration consists of a physical building floor and as well as an arrangement of APs within it (we use a combination of commodity 802.11ac APs manufactured by Netgear R6300v2 \cite{netgear:R6300v2} and ASUS RT-AC86U \cite{asus:RT-AC86U}). For simplicity, we set all APs in the environment to the same channel as the ESP32s. We carry out data collection inside and around the campus buildings. During data collection, we collect several traces by moving the sensors around while they are mounted on a pedestrian. This effectively simulates the trajectories of pedestrians entering the building. Furthermore, we also sample a set of test points in the building to assess location accuracy. For each test point, we collect around a few hundred Wi-Fi readings from the tandem ESP32s. We discard test points in which we detect less than 3 APs (in which case, the multi-lateration problem is under-determined). To compare \name{}'s performance to state-of-the-art baselines from both industry and academia, we also collect data for said baselines at the same test points.

\subsection{AP \& Client Location Accuracy}

\begin{figure*}
    \centering
    \begin{adjustbox}{valign=c}
    \begin{subfigure}[c]{0.33\linewidth}
        \centering
        \includegraphics[width=\linewidth]{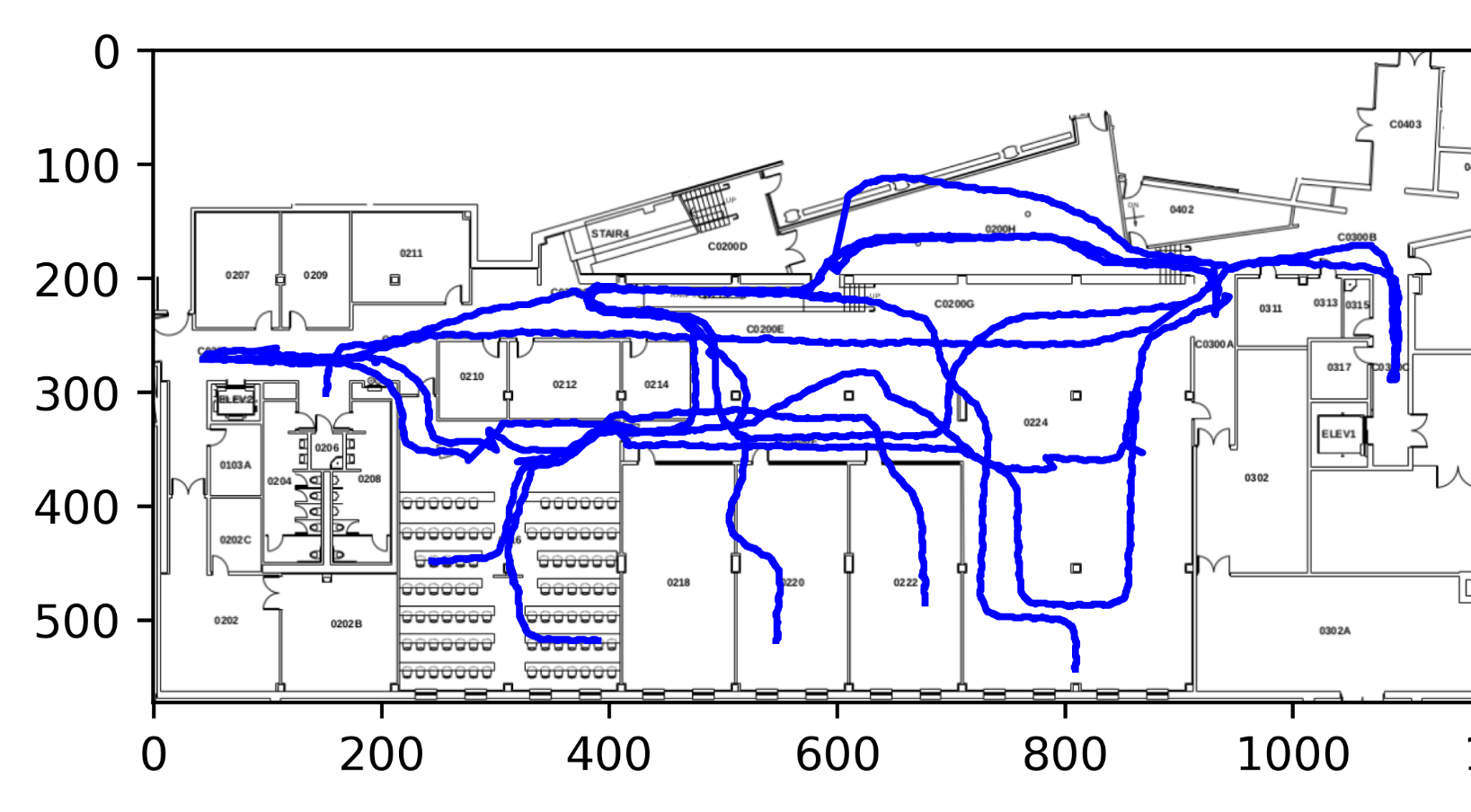}
        \caption{Pedestrian trajectories. }
        \label{fig:eval-example-a}
    \end{subfigure}
    \end{adjustbox}
    \hfill
    \begin{adjustbox}{valign=c}
    \begin{subfigure}[c]{0.33\linewidth}
        \centering
        \includegraphics[width=\linewidth]{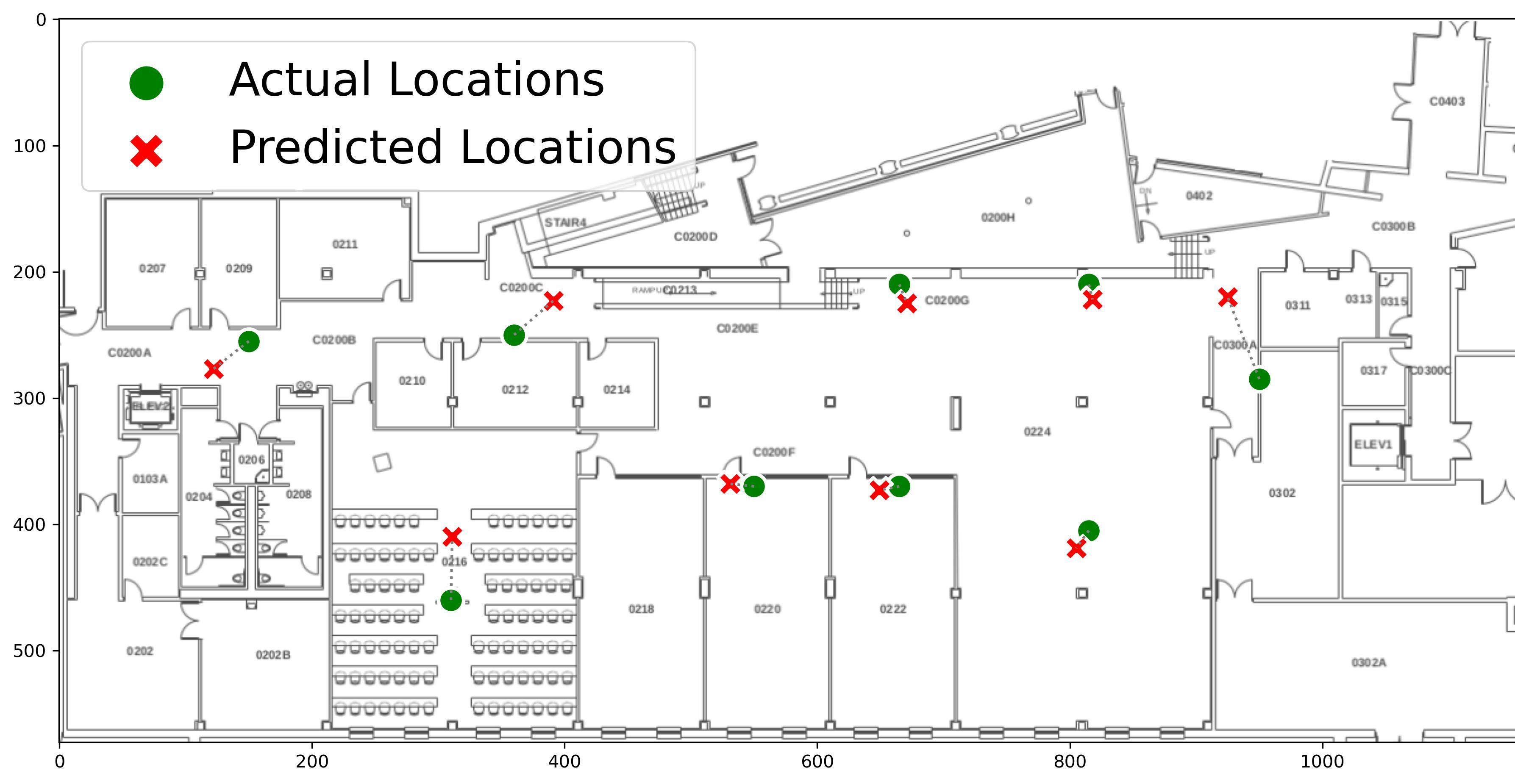}
        \caption{AP locations vs ground truth.}
        \label{fig:eval-example-b}
    \end{subfigure}
    \end{adjustbox}
    \hfill
    \begin{adjustbox}{valign=c}
    \begin{subfigure}[c]{0.33\linewidth}
        \centering
        \includegraphics[width=\linewidth]{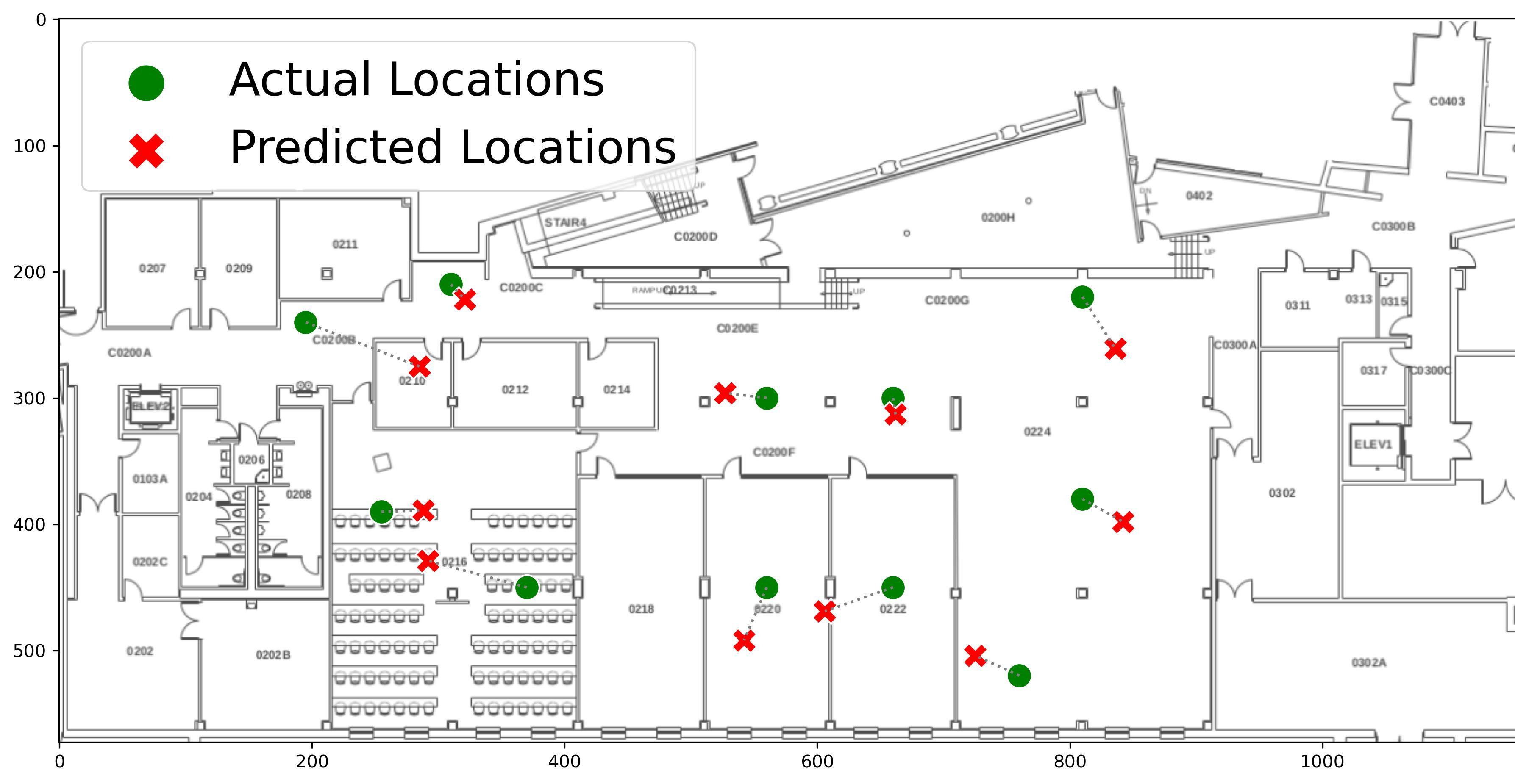}
        \caption{Client locations vs ground truth.}
        \label{fig:eval-example-c}
    \end{subfigure}
    \end{adjustbox}
    \caption{Sample \name{} run on floorplan A.}
    \label{fig:eval-example}
\end{figure*}

We show an example run of \name{} in Fig.~\ref{fig:eval-example}. In this environment, we collect 10 pedestrian trajectories as shown in Fig.~\ref{fig:eval-example-a}. Each trajectory is determined using an IMU dead-reckoning algorithm (RNIN-VIO \cite{chen_rnin-vio_2021}). For clarity, we truncate the beginning of each trajectory to originate at some entrance of the building, and is terminated after the pedestrian exceeds 70 m estimated travel distance (at which point we consider the dead-reckoning estimates no longer reliable). We run the algorithm to estimate the location of the APs and their processing times. The results of the algorithm are shown in Fig.~\ref{fig:eval-example-b}. Finally, we run the algorithm to localize the clients. The results are shown in Fig.~\ref{fig:eval-example-c}. 

\begin{figure*}[h!]
    \centering
    \begin{subfigure}[b]{0.32\linewidth}
        \centering
        \includegraphics[width=\linewidth]{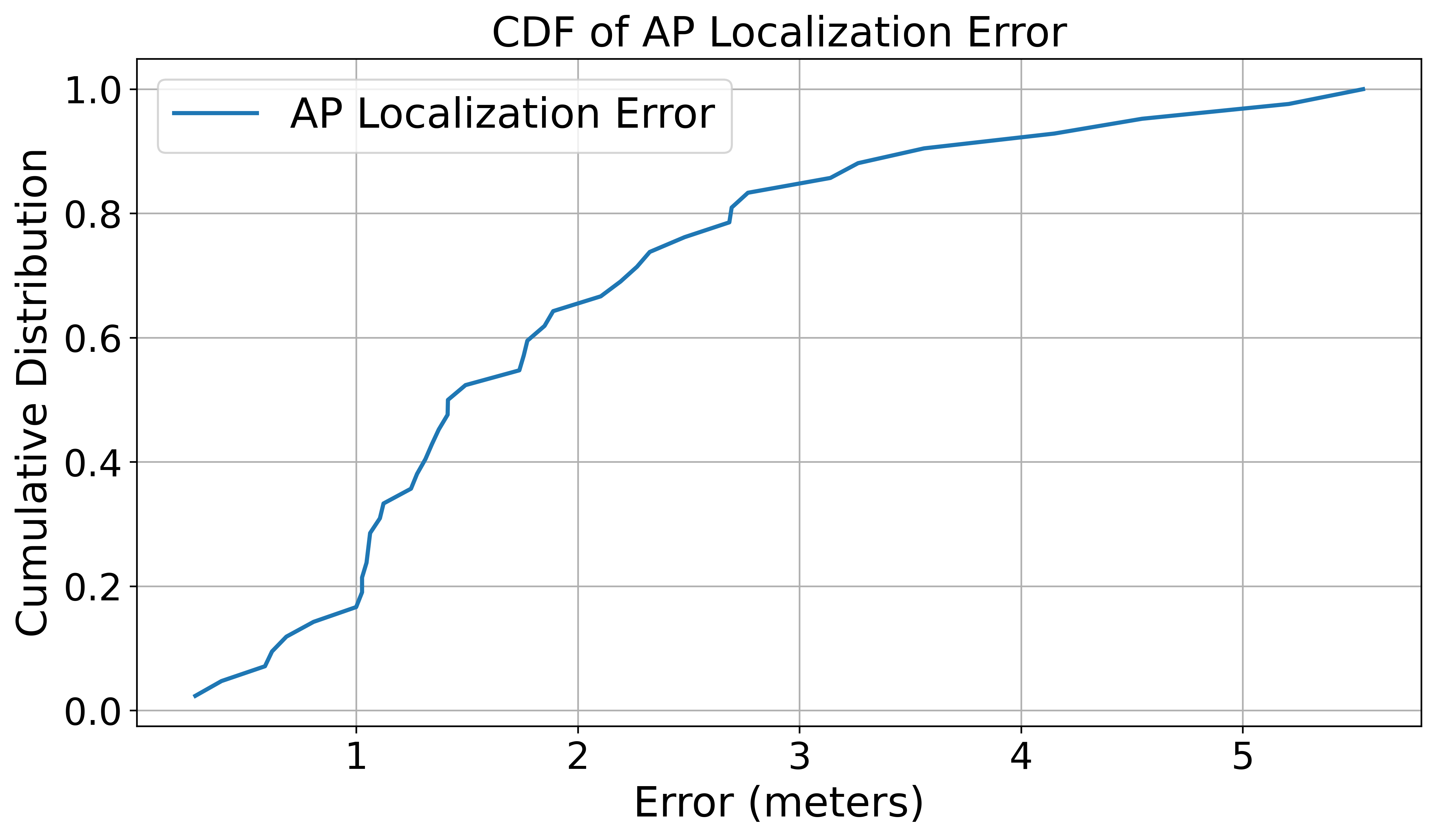}
        \caption{CDF of AP localization error.}
        \label{fig:eval-ap-cdf}
    \end{subfigure}
    \hfill
    \begin{subfigure}[b]{0.32\linewidth}
        \centering
        \includegraphics[width=\linewidth]{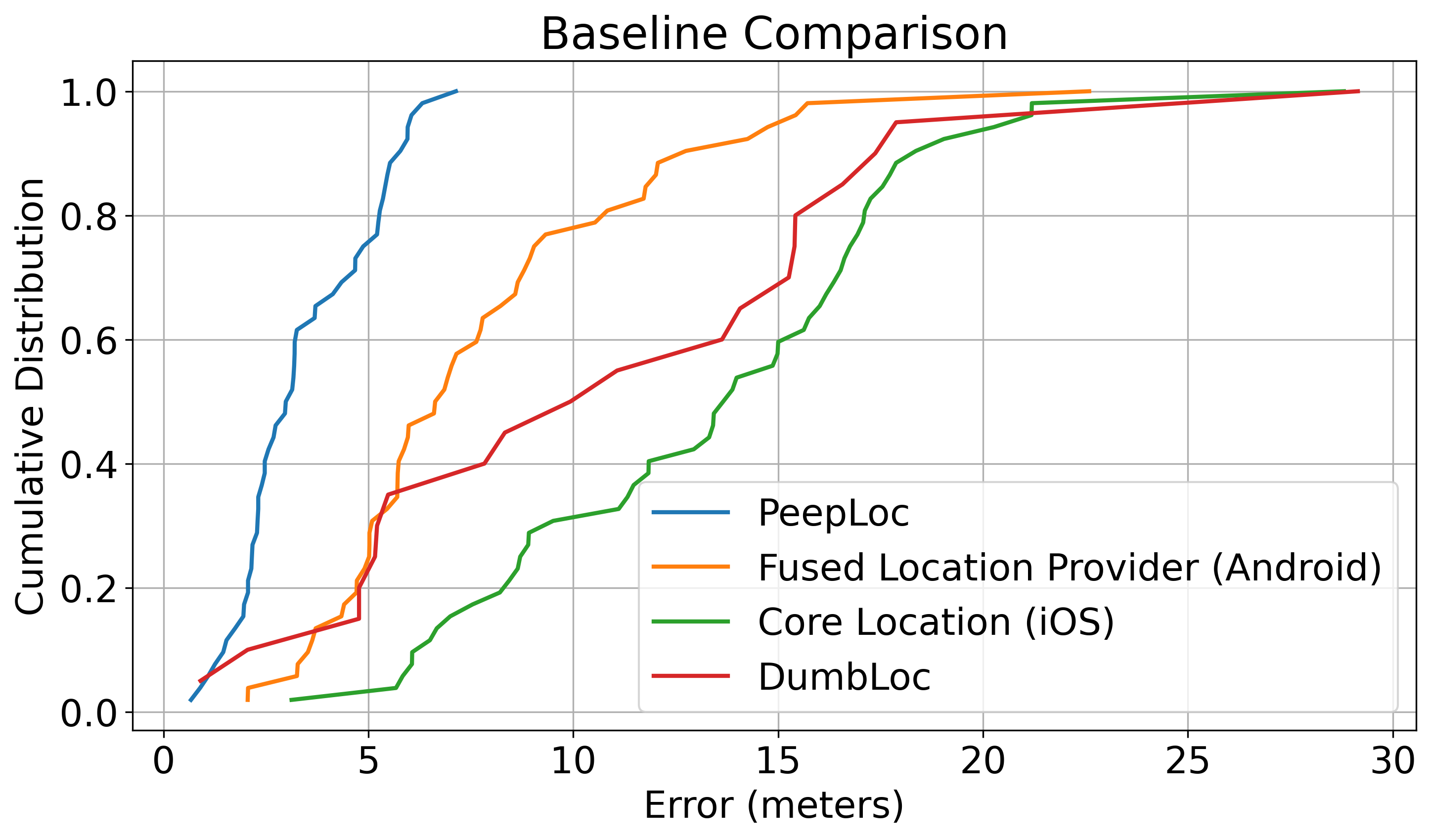}
        \caption{Comparison with baselines.}
        \label{fig:eval-results-cdf}
    \end{subfigure}
    \hfill
    \begin{subfigure}[b]{0.32\linewidth}
        \centering
        \includegraphics[width=\linewidth]{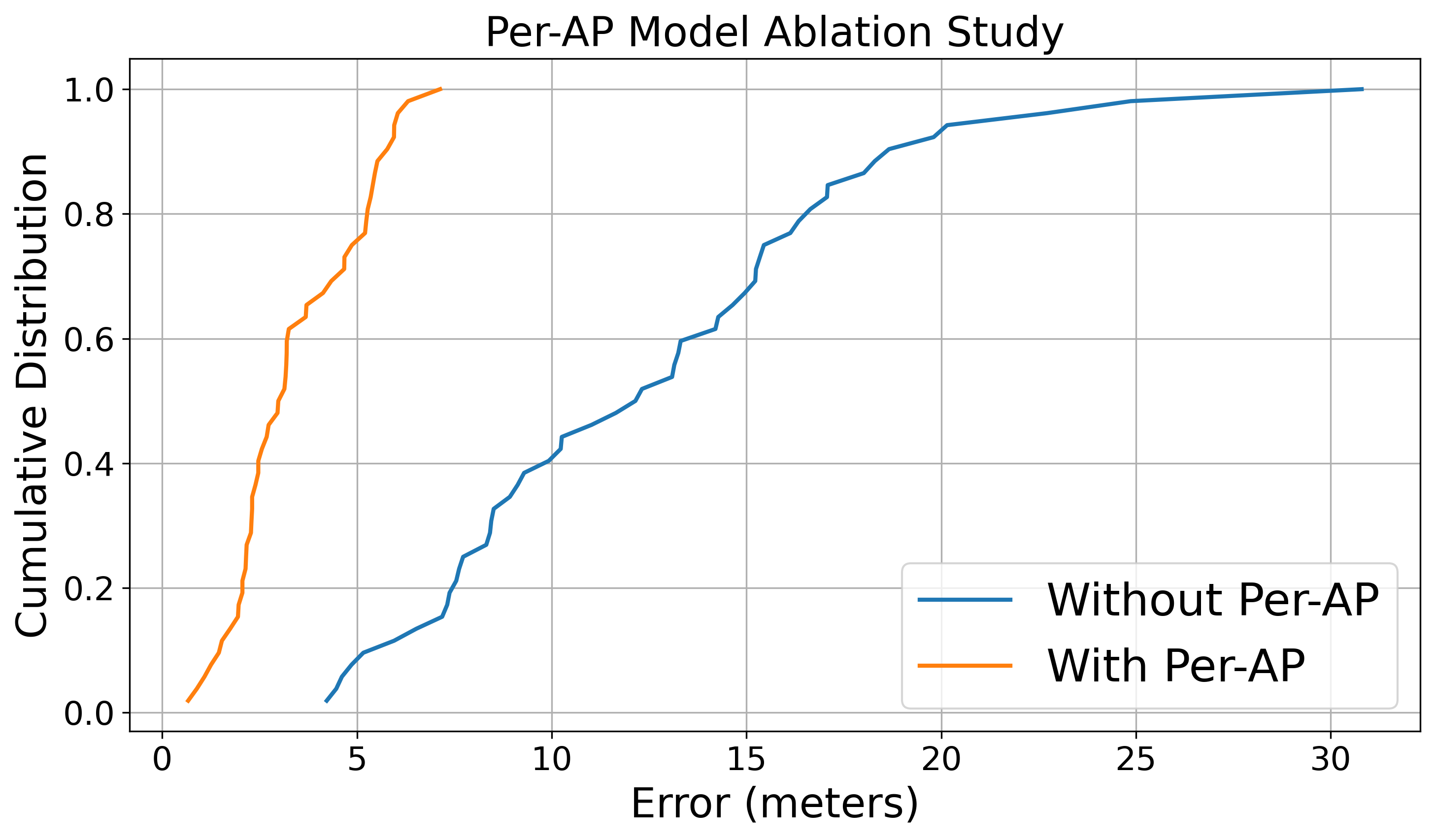}
        \caption{Ablation on per-AP ranging models.}
        \label{fig:ablation}
    \end{subfigure}

    \caption{Evaluation of \name{} across metrics: (a) AP localization accuracy, (b) comparison with baselines, and (c) effect of per-AP ranging models.}
    \label{fig:final-eval}
\end{figure*}

We repeat this process for all floorplans. To quantify \name{}'s AP localization performance, we show a CDF of all AP localization errors in Fig.~\ref{fig:eval-ap-cdf} -- the median error is 1.43 m. As the results show, \name{} is able to accurately reconstruct the arrangement of the APs across all environments. Similarly, we show a CDF of all client localization errors in Fig.~\ref{fig:eval-results-cdf}. We show that \name{} can localize clients with a mean accuracy of 3.41 m.

\subsection{Comparison with State-of-the-Art}

We compare \name{} with currently deployed indoor localization systems as well as state-of-the-art fingerprinting baselines.

\begin{itemize}
    \item \textbf{CoreLocation \cite{apple:corelocation}.} This is the geolocation service on iOS. This service uses Wi-Fi, GPS, Bluetooth, magnetometer, barometer, cellular hardware, and Bluetooth-based ranging measurements to nearby iBeacons. 
    \item \textbf{Fused Location Provider \cite{google:fusedlocationprovider}.} This is the Android geolocation service that combines various location sources, such as GPS, Wi-Fi, cellular networks, and ranging measurements towards nearby 802.11mc FTM-capable devices (e.g. Google Nest, Aruba APs). We note that all of the campus buildings that we run experiments in (i.e. all floorplans in Table.~\ref{tab:floorplans}) have an enterprise-grade deployment of FTM-capable Aruba AP-635 \cite{hpe_ap635_2025} running OpenLocate \cite{troymart_its_2023}.
    \item \textbf{DumbLoc~\cite{dumbloc_sensors2024}.} 
    A state-of-the-art machine learning-based Wi-Fi fingerprinting framework using XGBoost \cite{chen2016xgboost} and Random Forest models for indoor localization. It achieves meter-level accuracy through RSSI pattern matching without requiring parameter optimization or pre-mapped AP locations, designed for GPS-denied environments with existing Wi-Fi infrastructure. We configure DumbLoc to use the 5 APs with strongest signal strength for feature extraction.
\end{itemize}

We compare the performance of these services to \name{} in Fig.~\ref{fig:eval-results-cdf} and Table.~\ref{tab:eval-results}. The results show that \name{} is more accurate than all baselines. As expected, \name{} exceeds the performance of RSS-based DumbLoc \cite{dumbloc_sensors2024} due to relying on ranging measurements. The closest competitor to \name{} is the Fused Location Provider on Android, which synergizes with the GPS-located FTM-capable Aruba APs in the campus buildings. We attribute the poorer performance of the Fused Location Provider to (a) uncorrected device heterogeneity between our client and the Aruba APs and (b) errors in the locations of the APs as determined using OpenLocate. On the other hand, we attribute \name{}'s superior performance to its design which addresses both these challenges.

\begin{table*}[h!]
\centering
\caption{End-to-end client location accuracy statistics (in meters). Lower is better.}
\begin{tabular}{|l|c|c|c|}
\hline
\textbf{System} & \textbf{Mean (m)} & \textbf{Median (m)} & \textbf{Std. Deviation (m)} \\
\hline
Core Location (iOS) \cite{apple:corelocation}                    & 13.33 & 13.77 & 5.03 \\
DumbLoc~\cite{dumbloc_sensors2024}                               & 11.01 & 10.50 & 6.70 \\
Fused Location Provider (Android) \cite{google:fusedlocationprovider} + OpenLocate \cite{troymart_its_2023} & 7.71  & 6.74  & 3.98 \\
\rowcolor{LightGreen}
\textbf{\name{}}                                               & \textbf{3.41}  & \textbf{3.06}  & \textbf{1.63} \\
\hline
\end{tabular}
\label{tab:eval-results}
\end{table*}

%
%
%

\subsection{Ranging Model Ablation Study}

We conduct an ablation study on the per-AP ranging model by comparing the end-to-end client localization accuracy with and without the per-AP ranging model enabled (in which case we use a fixed slope of $\frac{c}{2}$). We summarize the results in Fig.~\ref{fig:ablation}. As the graph shows, the per-AP ranging model is crucial for accurate location estimation.

\subsection{Visualization} 
\begin{figure*}[t]
    \centering
    \begin{subfigure}[b]{0.4\textwidth}
        \centering
        \includegraphics[width=\linewidth]{figs/siebel_0_ap_pred_vs_gt_cropped.jpg}
        \caption{Floorplan A}
    \end{subfigure}
    \hspace{0.01\textwidth}
    \begin{subfigure}[b]{0.15\textwidth}
        \centering
        \includegraphics[width=\linewidth]{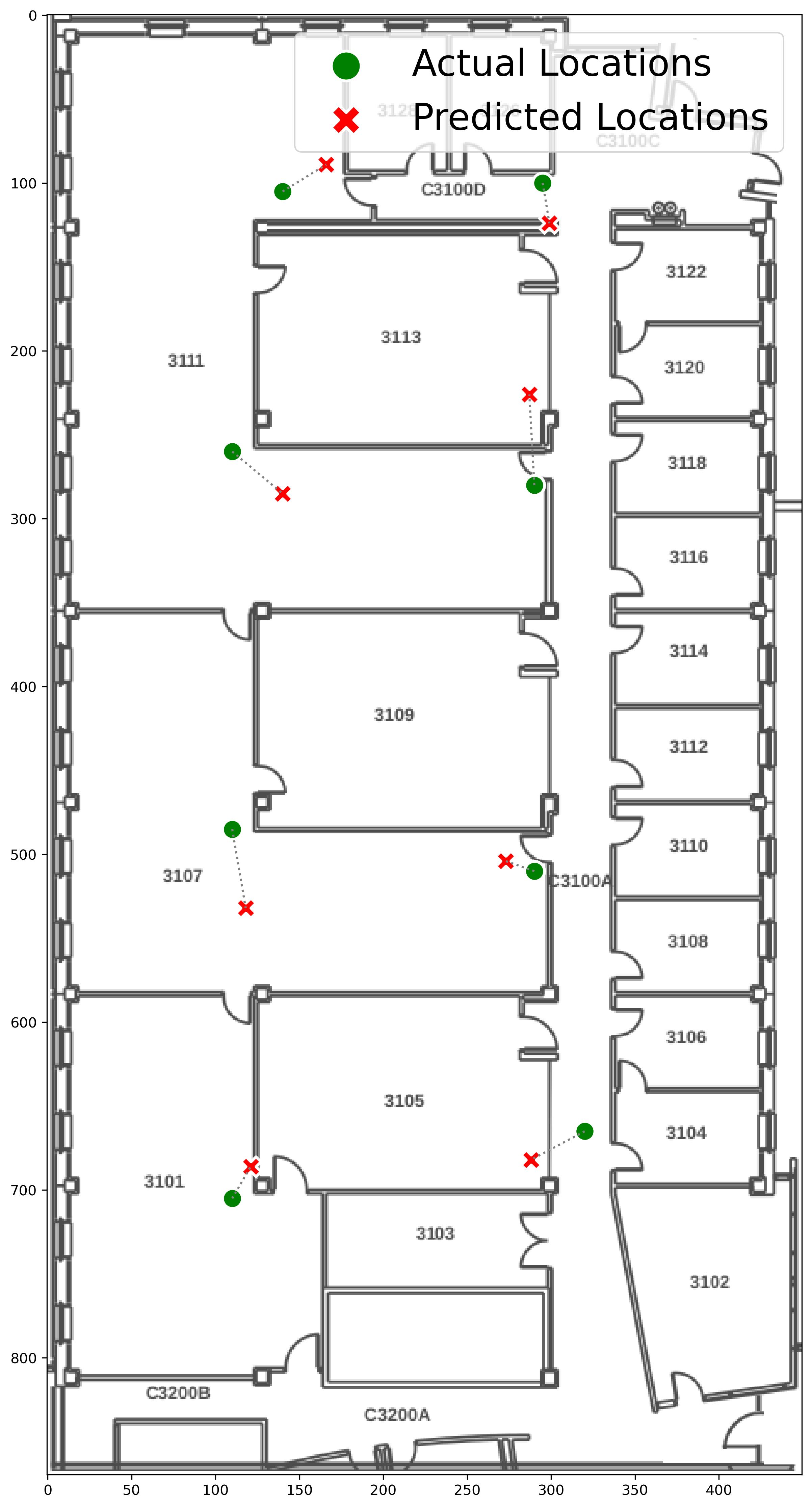}
        \caption{Floorplan B}
    \end{subfigure}

    \vspace{0.5em} 

    \begin{subfigure}[b]{0.2\textwidth}
        \centering
        \includegraphics[width=\linewidth]{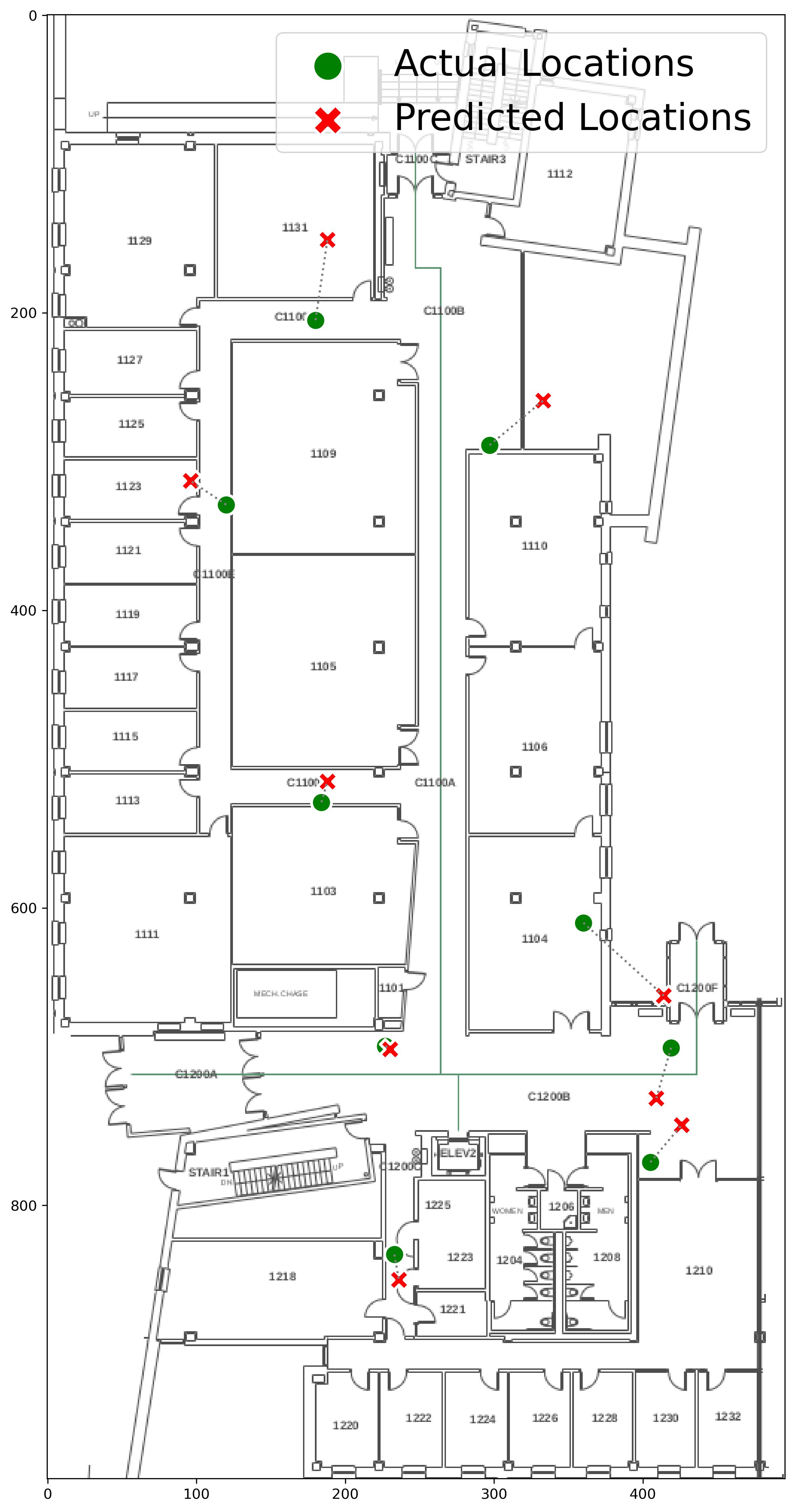}
        \caption{Floorplan C}
    \end{subfigure}
    \hspace{0.01\textwidth}
    \begin{subfigure}[b]{0.155\textwidth}
        \centering
        \includegraphics[width=\linewidth]{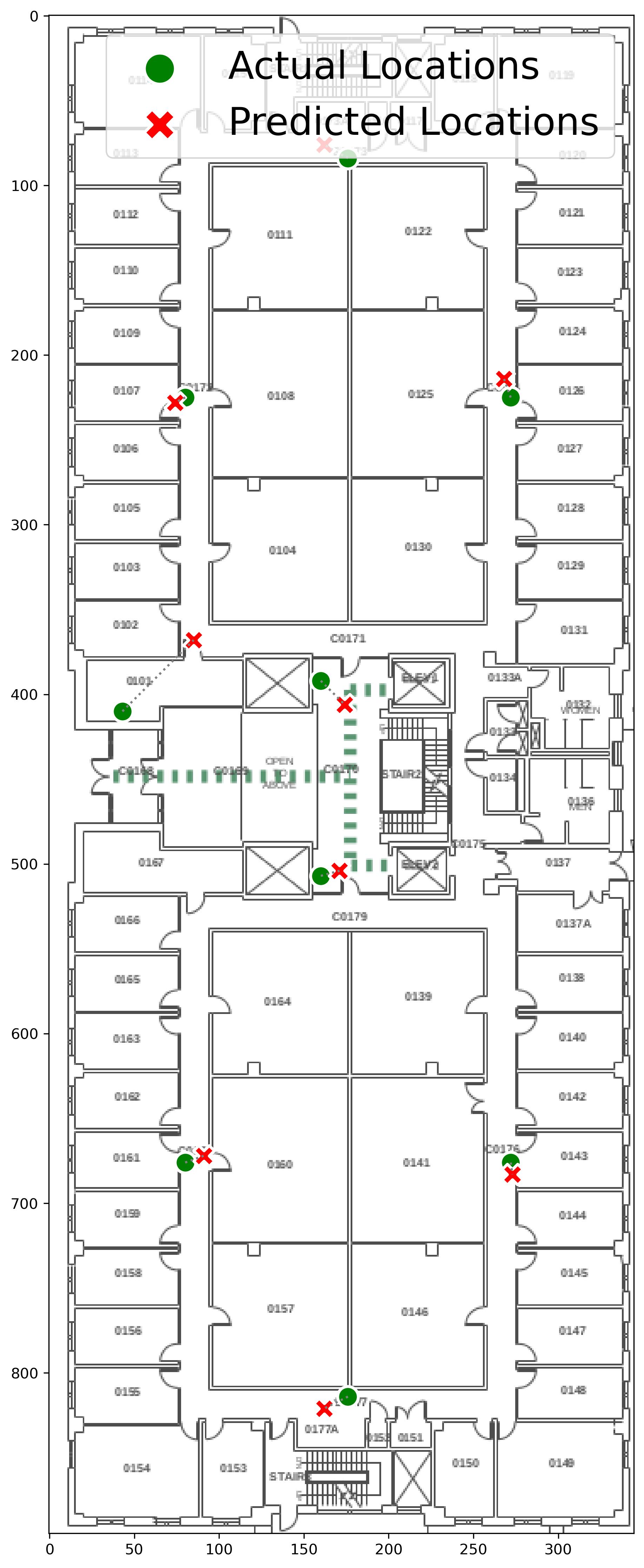}
        \caption{Floorplan D1}
    \end{subfigure}
    \hspace{0.01\textwidth}
    \begin{subfigure}[b]{0.155\textwidth}
        \centering
        \includegraphics[width=\linewidth]{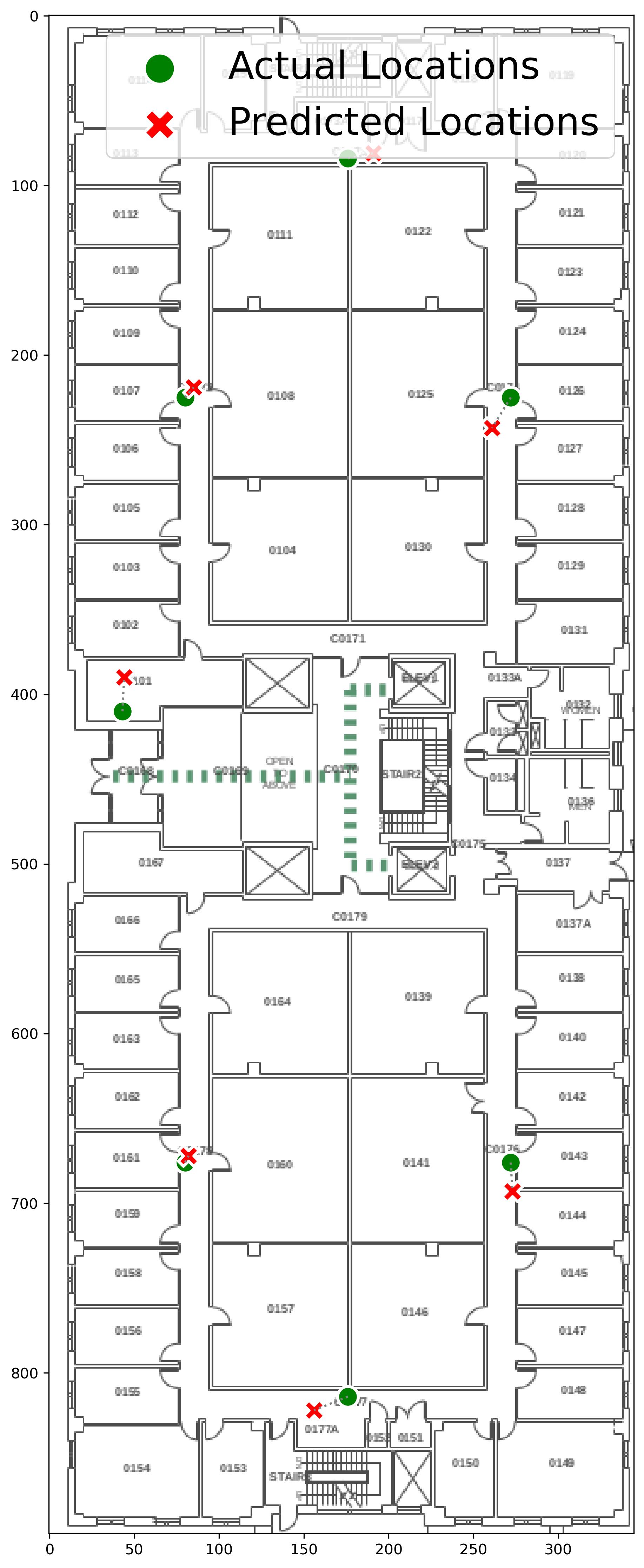}
        \caption{Floorplan D2}
    \end{subfigure}

    \caption{Building floorplans with predicted and ground truth APs marked. Best viewed zoomed-in.}
    \label{fig:ap_pred_vs_gt}
\end{figure*}

\begin{figure*}[t]
    \centering
    \begin{subfigure}[b]{0.40\textwidth}
        \centering
        \includegraphics[width=\linewidth]{figs/siebel_0_client_pred_vs_gt_cropped.jpg}
        \caption{Floorplan A}
    \end{subfigure}
    \hspace{0.01\textwidth}
    \begin{subfigure}[b]{0.15\textwidth}
        \centering
        \includegraphics[width=\linewidth]{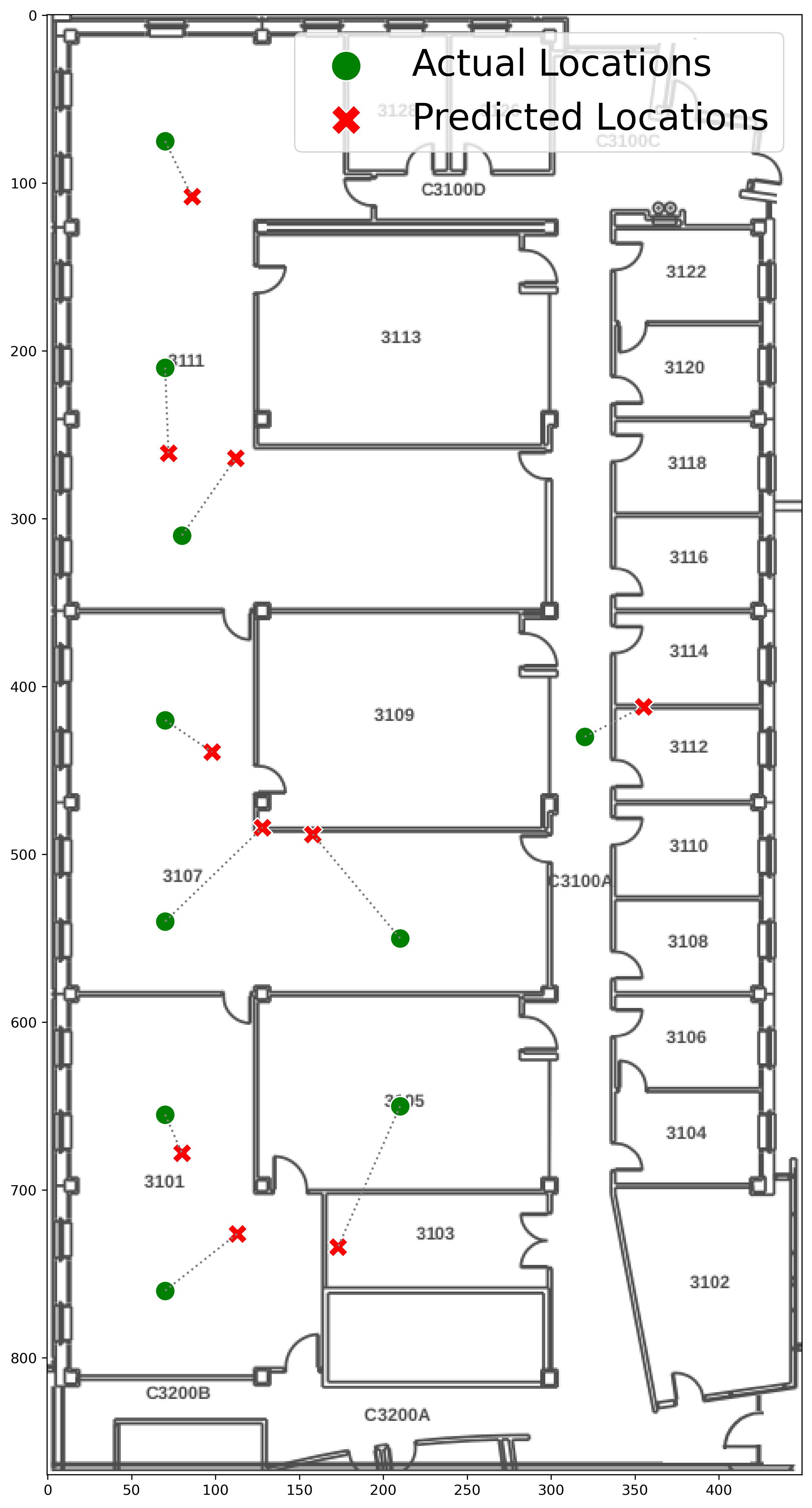}
        \caption{Floorplan B}
    \end{subfigure}

    \vspace{0.5em} 

    \begin{subfigure}[b]{0.2\textwidth}
        \centering
        \includegraphics[width=\linewidth]{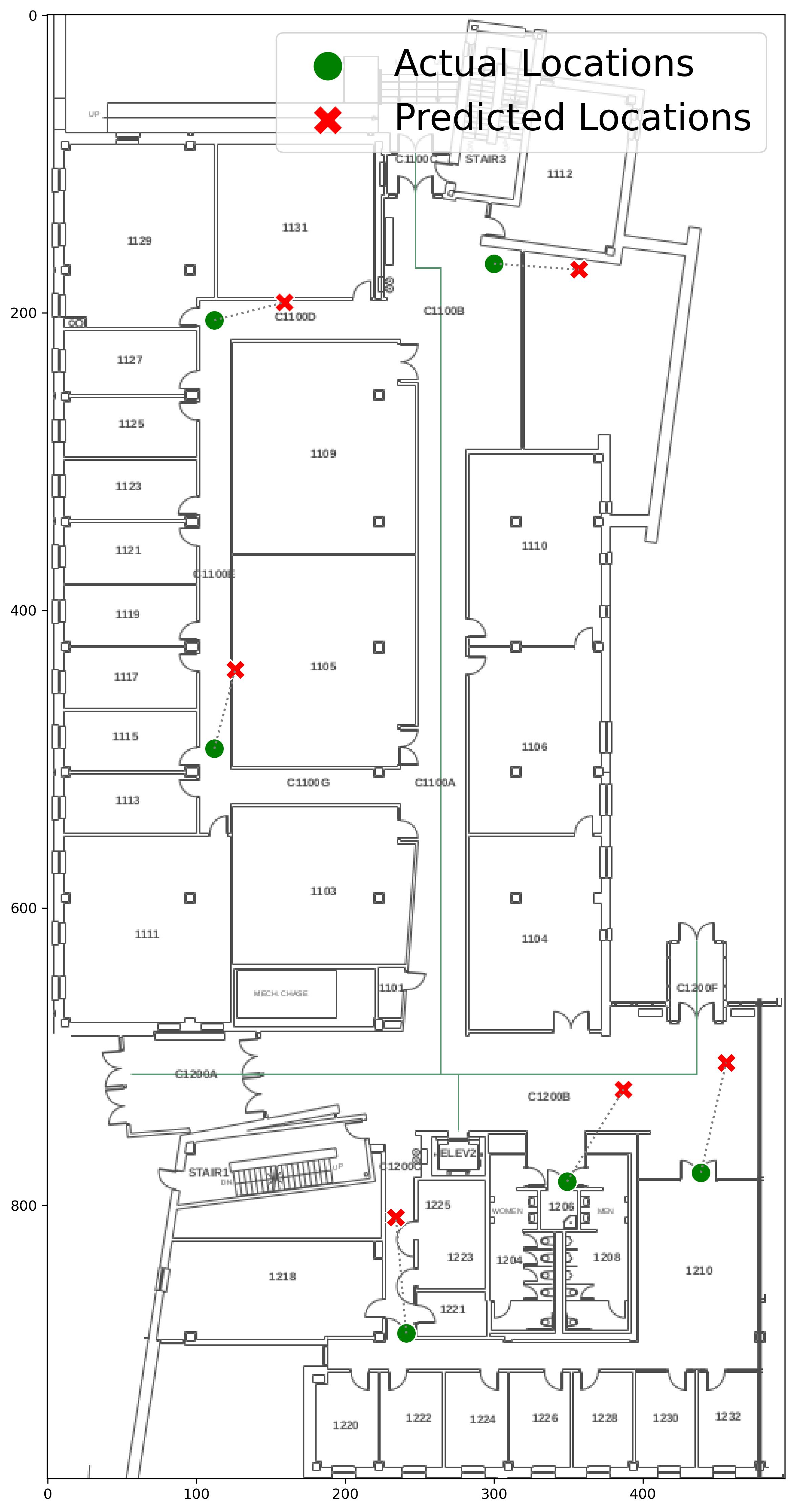}
        \caption{Floorplan C}
    \end{subfigure}
    \hspace{0.01\textwidth}
    \begin{subfigure}[b]{0.155\textwidth}
        \centering
        \includegraphics[width=\linewidth]{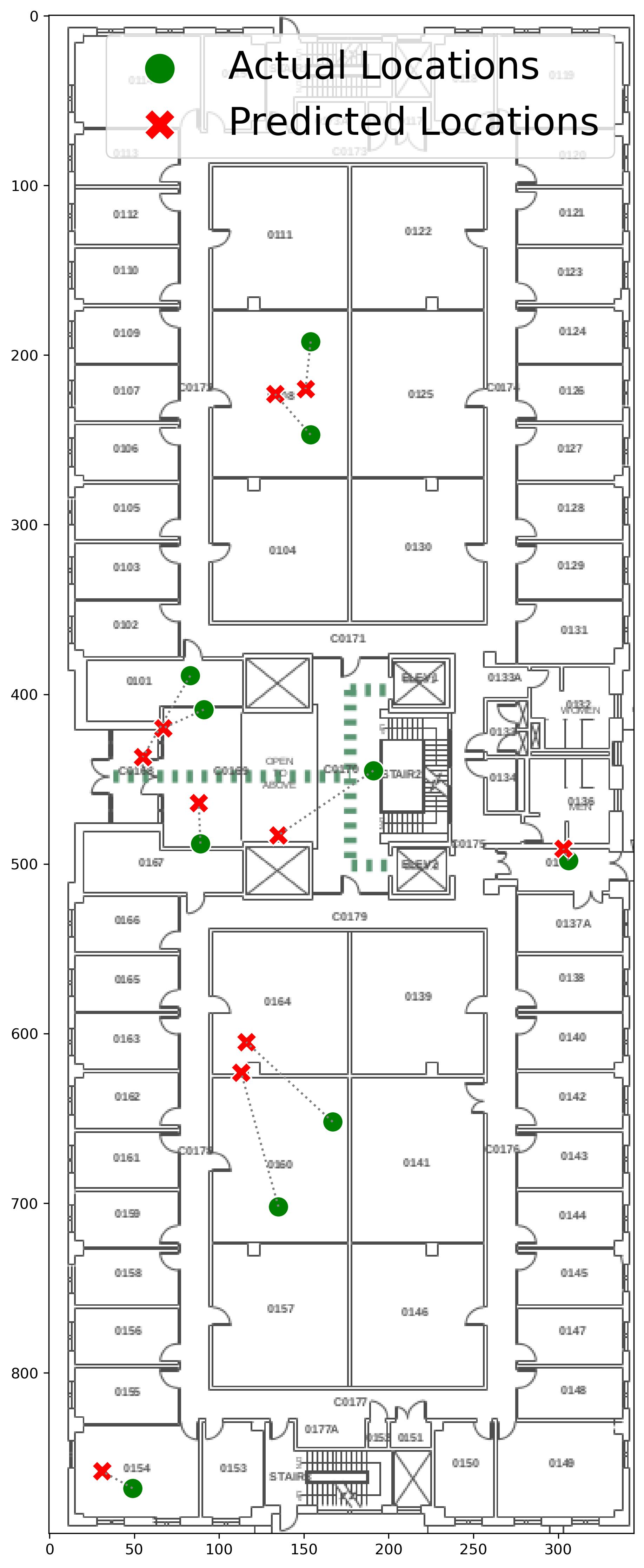}
        \caption{Floorplan D1}
    \end{subfigure}
    \hspace{0.01\textwidth}
    \begin{subfigure}[b]{0.155\textwidth}
        \centering
        \includegraphics[width=\linewidth]{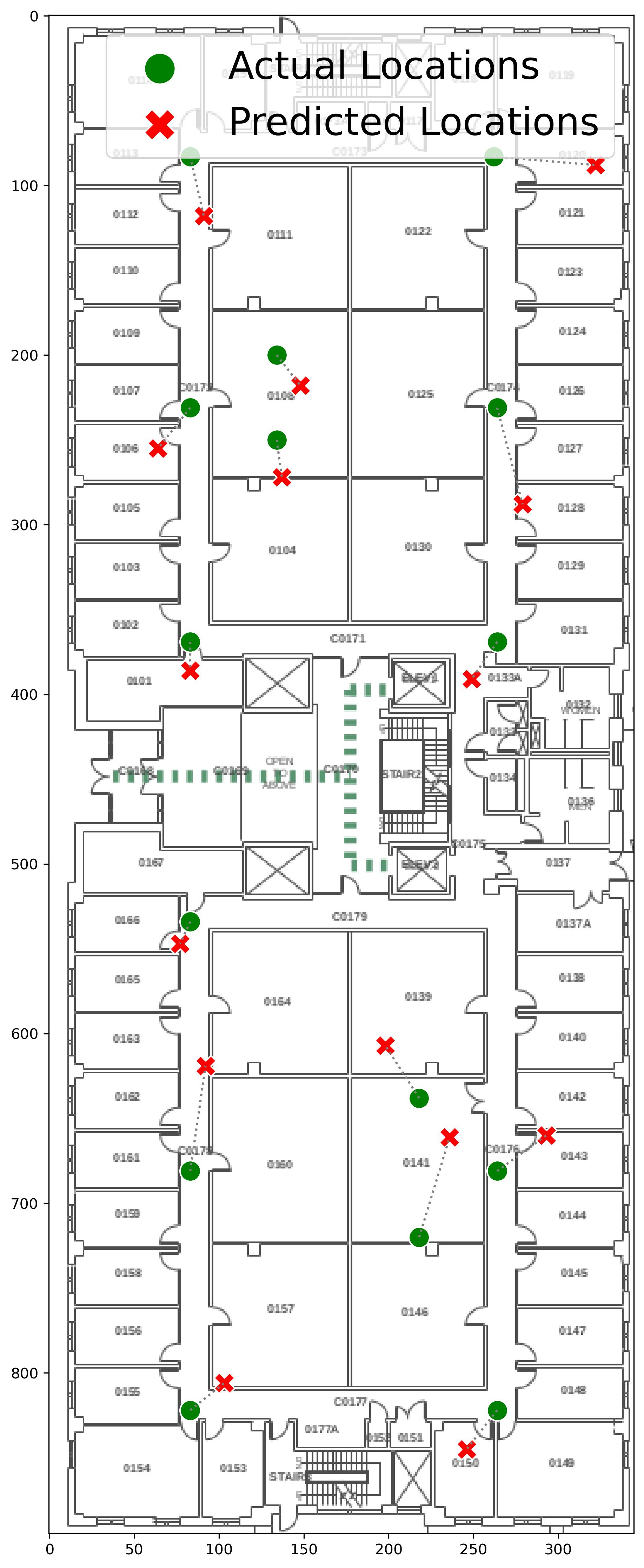}
        \caption{Floorplan D2}
    \end{subfigure}

    \caption{Building floorplans with predicted and ground truth clients marked. Best viewed zoomed-in.}
    \label{fig:client_pred_vs_gt}
\end{figure*}

For completeness, we show the output of \name{}'s AP localization algorithm in Fig.~\ref{fig:ap_pred_vs_gt} and \name{}'s client localization algorithm in Fig.~\ref{fig:client_pred_vs_gt} respectively. For ease of reference, the results are superimposed onto the building floorplans.

\section{Discussion}
\label{sec:discussion}

\para{Mapping Larger Environments.} Can \name{} can scale to map larger buildings than those considered in this work? We believe the answer is yes. Just as how \name{} leverages dead reckoning from a known position provided by GPS, \name{} can also leverage dead reckoning from a known position provided by \name{} itself. This means that larger buildings can be mapped incrementally, first starting from regions in the outer edges, and then gradually expanding inwards until the entire building is mapped out. 

\para{Handling Multiple Floors.} In this paper, we map only the single floors of buildings considered in this work. One natural question to ask is whether \name{} can scale to multi-floor buildings. We believe there are two potential avenues to make this work. First, we can simply extend \name{} to work in 3D rather than in a 2D planar environment. This should be possible since ToF measurements are done in 3D. Second, we can break down a building into separate floors, and apply \name{} to each individual floor. We leave such investigation to future work.

\para{Change Detection \& Handling.} What happens if the location of the Wi-Fi infrastructure in a building changes? This is a natural question to ask since we can reasonably expect Wi-Fi APs in a building to change their location within some time frame due to being moved around. We believe that such changes can be detected using inertial dead reckoning. Specifically, when the known position of a pedestrian from inertial dead reckoning and the estimated location from \name{} diverge, this tells us that the arrangement of the APs in the environment has changed. When this change is detected, \name{} can re-run the AP discovery and localization phase and reinitialize the system. Hence, \name{} can be deployed as a self-correcting system.

\para{Integration with Next-Gen Systems.} The recently proposed 802.11az standard, also known as Next Generation Positioning (NGP) \cite{IEEE802.11az2022}, purports to further enhance the Wi-Fi ranging capabilities introduced by IEEE 802.11mc (Fine Timing Measurement, FTM). Specifically, it adds wider channel bandwidths (up to 160 MHz with Wi-Fi 6 and potentially 320 MHz with Wi-Fi 7) for more fine-grained ranging resolution, and employs Multiple Input Multiple Output (MIMO) techniques to mitigate multipath effects. We note that \name{} can benefit from and extend the widespread deployment of such 802.11az-capable APs. Specifically, when 802.11az APs are deployed in an environment, their locations can be inferred or refined using \name{}'s crowdsourcing model. 

\section{Conclusion}
\label{sec:conclusion}

We present \name{}, a Wi-Fi ranging-based indoor positioning system that can be ubiquitously deployed. \name{} requires no additional hardware beyond that present in commodity smartphones, and no infrastructure requirements beyond pre-existing Wi-Fi APs. Unlike existing indoor localization systems, \name{} requires no cooperation from the infrastructure, and can bootstrap itself from pedestrian data with no dedicated human effort. In this paper, we show how \name{} can be deployed in various campus buildings, showing improved performance over both fingerprinting-based approaches as well as state-of-the-art Wi-Fi ranging-based approaches (e.g. FTM). Future work can extend \name{} to work on buildings with multiple floors, or handle changes in the arrangement of Wi-Fi APs over time.

{\footnotesize \bibliographystyle{acm}
\bibliography{references}}

\begin{thebibliography}{10}

\bibitem{stereolabs_global_2025}
Global {Localization} {Overview} - {Stereolabs}.

\bibitem{hpe_ap635_2025}
{HPE} {Aruba} {Networking} 630 {Series} {Campus} {Access} {Points}.

\bibitem{apple_indoor_maps}
Introducing the {Indoor} {Maps} {Program}.

\bibitem{IEEE802.11mc2016}
{IEEE} standard for information technology—telecommunications and information exchange between systems—local and metropolitan area networks—specific requirements part 11: Wireless {LAN} medium access control (mac) and physical layer (phy) specifications, May 2016.
\newblock Maintenance revision by Task Group mc (TGmc), sometimes referred to as IEEE 802.11mc.

\bibitem{IEEE802.11az2022}
{IEEE} draft standard for information technology—telecommunications and information exchange between systems—local and metropolitan area networks—specific requirements part 11: Wireless {LAN} medium access control (mac) and physical layer (phy) specifications—amendment 4: Enhancements for positioning, Mar. 2023.
\newblock Published as the next-generation positioning (NGP) amendment.

\bibitem{abedi_non-cooperative_2022}
{\sc Abedi, A., and Vasisht, D.}
\newblock Non-cooperative wi-fi localization \& its privacy implications.
\newblock In {\em Proceedings of the 28th {Annual} {International} {Conference} on {Mobile} {Computing} {And} {Networking}\/} (Sydney NSW Australia, Oct. 2022), ACM, pp.~570--582.

\bibitem{aggarwal_is_2022}
{\sc Aggarwal, S., Sheshadri, R.~K., Sundaresan, K., and Koutsonikolas, D.}
\newblock Is wifi 802.11mc fine time measurement ready for prime-time localization?
\newblock In {\em Proceedings of the 16th {ACM} {Workshop} on {Wireless} {Network} {Testbeds}, {Experimental} evaluation \& {CHaracterization}\/} (Sydney NSW Australia, Oct. 2022), ACM, pp.~1--8.

\bibitem{apple:corelocation}
{\sc {Apple Inc.}}
\newblock {\em Core Location Framework Reference}.
\newblock Apple Developer, Cupertino, CA, USA, 2025.
\newblock Provides services for geographic location, altitude, orientation, and region monitoring.

\bibitem{arun_viwid_2022}
{\sc Arun, A., Hunter, W., Ayyalasomayajula, R., and Bharadia, D.}
\newblock {ViWiD}: {Leveraging} {WiFi} for {Robust} and {Resource}-{Efficient} {SLAM}, Sept. 2022.
\newblock arXiv:2209.08091 [cs].

\bibitem{asus:RT-AC86U}
{\sc {ASUSTeK Computer Inc.}}
\newblock {\em ASUS RT-AC86U Dual-band Gigabit WiFi Router Data Sheet}.
\newblock ASUS, Taipei, Taiwan, 2017.
\newblock 1.8 GHz dual-core CPU; 3×3 (2.4 GHz) + 4×4 (5 GHz) MIMO (AC2900); MU‑MIMO; RangeBoost.

\bibitem{ayyalasomayajula_locap_nodate}
{\sc Ayyalasomayajula, R., Arun, A., Wu, C., Rajagopalan, S., Ganesaraman, S., Seetharaman, A., Jain, I.~K., and Bharadia, D.}
\newblock {LocAP}: {Autonomous} {Millimeter} {Accurate} {Mapping} of {WiFi} {Infrastructure}.

\bibitem{bahl2000radar}
{\sc Bahl, P., and Padmanabhan, V.~N.}
\newblock Radar: An in-building rf-based user location and tracking system.
\newblock In {\em Proceedings IEEE INFOCOM 2000\/} (2000), vol.~2, IEEE, pp.~775--784.

\bibitem{chen_ionet_2018}
{\sc Chen, C., Lu, X., Markham, A., and Trigoni, N.}
\newblock {IONet}: {Learning} to {Cure} the {Curse} of {Drift} in {Inertial} {Odometry}.
\newblock {\em Proceedings of the AAAI Conference on Artificial Intelligence 32}, 1 (Apr. 2018).

\bibitem{chen_deep_2024}
{\sc Chen, C., and Pan, X.}
\newblock Deep {Learning} for {Inertial} {Positioning}: {A} {Survey}, Mar. 2024.
\newblock arXiv:2303.03757 [cs].

\bibitem{chen_rnin-vio_2021}
{\sc Chen, D., Wang, N., Xu, R., Xie, W., Bao, H., and Zhang, G.}
\newblock {RNIN}-{VIO}: {Robust} {Neural} {Inertial} {Navigation} {Aided} {Visual}-{Inertial} {Odometry} in {Challenging} {Scenes}.
\newblock In {\em 2021 {IEEE} {International} {Symposium} on {Mixed} and {Augmented} {Reality} ({ISMAR})\/} (Bari, Italy, Oct. 2021), IEEE, pp.~275--283.

\bibitem{chen2016xgboost}
{\sc Chen, T., and Guestrin, C.}
\newblock Xgboost: A scalable tree boosting system.
\newblock In {\em Proceedings of the 22nd {ACM} {SIGKDD} International Conference on Knowledge Discovery and Data Mining (KDD ’16)\/} (2016), ACM, pp.~785--794.

\bibitem{google:fusedlocationprovider}
{\sc {Google Inc.}}
\newblock {\em FusedLocationProviderClient — Google Play Services Location API}.
\newblock Google, Mountain View, CA, USA, 2025.
\newblock High-level, battery-efficient location provider combining GPS, Wi‑Fi, and cellular data.

\bibitem{herath_ronin_2020}
{\sc Herath, S., Yan, H., and Furukawa, Y.}
\newblock {RoNIN}: {Robust} {Neural} {Inertial} {Navigation} in the {Wild}: {Benchmark}, {Evaluations}, \& {New} {Methods}.
\newblock {\em 2020 IEEE International Conference on Robotics and Automation (ICRA)\/} (May 2020), 3146--3152.
\newblock Conference Name: 2020 IEEE International Conference on Robotics and Automation (ICRA) ISBN: 9781728173955 Place: Paris, France Publisher: IEEE.

\bibitem{horn_indoor_2022}
{\sc Horn, B. K.~P.}
\newblock Indoor {Localization} {Using} {Uncooperative} {Wi}-{Fi} {Access} {Points}.
\newblock {\em Sensors 22}, 8 (Jan. 2022), 3091.
\newblock Number: 8 Publisher: Multidisciplinary Digital Publishing Institute.

\bibitem{ibrahim2018rtt}
{\sc Ibrahim, M., He, S., Liu, C., Cheng, B., Gruteser, M., Chen, Y., and Martin, R.~P.}
\newblock Accurate wifi-based localization for smartphones using peer-to-peer fine time measurements.
\newblock In {\em Proceedings of the 24th Annual International Conference on Mobile Computing and Networking\/} (2018), pp.~1--13.

\bibitem{kotaru2015spotfi}
{\sc Kotaru, M., Joshi, K., Bharadia, D., and Katti, S.}
\newblock Spotfi: Decimeter level localization using wifi.
\newblock In {\em Proceedings of the 2015 ACM Conference on Special Interest Group on Data Communication\/} (2015), pp.~269--282.

\bibitem{liu_tlio_2020}
{\sc Liu, W., Caruso, D., Ilg, E., Dong, J., Mourikis, A.~I., Daniilidis, K., Kumar, V., and Engel, J.}
\newblock {TLIO}: {Tight} {Learned} {Inertial} {Odometry}.
\newblock {\em IEEE Robotics and Automation Letters 5}, 4 (Oct. 2020), 5653--5660.

\bibitem{dumbloc_sensors2024}
{\sc Narasimman, S.~C., and Alphones, A.}
\newblock Dumbloc: Dumb indoor localization framework using wi-fi fingerprinting.
\newblock {\em IEEE Sensors Journal 24}, 9 (2024), 14623 – 14630.

\bibitem{netgear:R6300v2}
{\sc {NETGEAR, Inc.}}
\newblock {\em NETGEAR R6300 Smart WiFi Router (AC1750) User Manual}.
\newblock NETGEAR, San Jose, CA, USA, 2016.
\newblock Dual-core 800 MHz CPU; 3x3 (2.4 GHz) + 3x3 (5 GHz) MIMO; 1 × USB 3.0; Beamforming+ AC1750.

\bibitem{ni_experience_2022}
{\sc Ni, J., Zhang, F., Xiong, J., Huang, Q., Chang, Z., Ma, J., Xie, B., Wang, P., Bian, G., Li, X., and Liu, C.}
\newblock Experience: pushing indoor localization from laboratory to the wild.
\newblock In {\em Proceedings of the 28th {Annual} {International} {Conference} on {Mobile} {Computing} {And} {Networking}\/} (Sydney NSW Australia, Oct. 2022), ACM, pp.~147--157.

\bibitem{nowicki2017wifi}
{\sc Nowicki, M., and Wietrzykowski, J.}
\newblock Wifi-lstm: Recurrent neural network-based indoor localization using csi time series.
\newblock In {\em Proceedings of the International Conference on Indoor Positioning and Indoor Navigation (IPIN)\/} (2017), IEEE.

\bibitem{chintalapudi_zee_2012}
{\sc Rai, A., Chintalapudi, K.~K., Padmanabhan, V.~N., and Sen, R.}
\newblock Zee: zero-effort crowdsourcing for indoor localization.
\newblock In {\em Proceedings of the 18th Annual International Conference on Mobile Computing and Networking\/} (New York, NY, USA, 2012), Mobicom '12, Association for Computing Machinery, p.~293–304.

\bibitem{rye_surveilling_2024}
{\sc Rye, E., and Levin, D.}
\newblock Surveilling the {Masses} with {Wi}-{Fi}-{Based} {Positioning} {Systems}.
\newblock In {\em 2024 {IEEE} {Symposium} on {Security} and {Privacy} ({SP})\/} (San Francisco, CA, USA, May 2024), IEEE, pp.~2831--2846.

\bibitem{stereolabs_zed_2025}
{\sc Stereolabs}.
\newblock {ZED} 2i {\textbar} {Stereo} {Camera} {\textbar} {Stereolabs}.

\bibitem{espressif_esp32_s3_2025}
{\sc Systems, E.}
\newblock {ESP32}-{S3} {Wi}-{Fi} \& {BLE} 5 {SoC} {\textbar} {Espressif} {Systems}.

\bibitem{troymart_its_2023}
{\sc troymart}.
\newblock It’s been a year, where is {Open} {Locate} {Now}? - {Wi}-{Fi} {Vitae}, June 2023.

\bibitem{ublox_zed_f9p_2023}
{\sc u~blox}.
\newblock {ZED}-{F9P} module, Apr. 2023.

\bibitem{vasisht2016chronos}
{\sc Vasisht, D., Kumar, S., and Katabi, D.}
\newblock Chronos: Sub-nanosecond time-of-flight measurement using commodity wifi cards.
\newblock In {\em Proceedings of the 2016 Conference on Special Interest Group on Data Communication\/} (2016), pp.~105--117.

\bibitem{wang_no_2012}
{\sc Wang, H., Sen, S., Elgohary, A., Farid, M., Youssef, M., and Choudhury, R.~R.}
\newblock No need to war-drive: unsupervised indoor localization.
\newblock In {\em Proceedings of the 10th international conference on {Mobile} systems, applications, and services\/} (New York, NY, USA, June 2012), {MobiSys} '12, Association for Computing Machinery, pp.~197--210.

\bibitem{xiong_arraytrack_2013}
{\sc Xiong, J., and Jamieson, K.}
\newblock Arraytrack: a fine-grained indoor location system.
\newblock In {\em Proceedings of the 10th USENIX Conference on Networked Systems Design and Implementation\/} (USA, 2013), nsdi'13, USENIX Association, p.~71–84.

\bibitem{youssef2005horus}
{\sc Youssef, M., and Agrawala, A.}
\newblock The horus wlan location determination system.
\newblock In {\em Proceedings of the 3rd international conference on Mobile systems, applications, and services\/} (2005), pp.~205--218.

\end{thebibliography}


\end{document}